\documentclass[sn-aps]{sn-jnl}

\usepackage{graphicx}%
\usepackage{multirow}%
\usepackage{amsmath,amssymb,amsfonts}%
\usepackage{amsthm}%
\usepackage{mathrsfs}%
\usepackage[title]{appendix}%
\usepackage{xcolor}%
\usepackage{textcomp}%
\usepackage{manyfoot}%
\usepackage{booktabs}%
\usepackage{algorithm}%
\usepackage{algorithmicx}%
\usepackage{algpseudocode}%
\usepackage{listings}%

\theoremstyle{thmstyleone}%
\theoremstyle{thmstyletwo}%

\theoremstyle{thmstylethree}%

\begin{document}
\title[Article Title]{Constraining the $f$-mode oscillations frequency in Neutron Stars through Universal Relations in the realm of  Energy-Momentum Squared Gravity}


\author[]{\fnm{Sayantan} \sur{Ghosh}}\email{sayantanghosh1999@gmail.com}



\affil[]{\orgdiv{Department of Physics and Astronomy}, \orgname{National Institute of Technology Rourkela}, \orgaddress{\city{Rourkela}, \postcode{769008}, \state{Odisha}, \country{India}}}




\abstract{

Neutron stars (NSs), superdense objects with exceptionally strong gravitational fields, provide an ideal laboratory for probing general relativity (GR) in the high-curvature regime. They also present an exciting opportunity to explore new gravitational physics beyond the traditional framework of GR. Thus, investigating alternative theories of gravity in the context of superdense stars is intriguing and essential for advancing our understanding of gravitational phenomena in extreme environments.
Energy-Momentum Squared Gravity (EMSG) is a modified theory of gravity that extends GR by including nonlinear terms involving the energy-momentum tensor $T_{\mu \nu}$. This study examines the effects of EMSG on the properties and behaviour of NSs by varying the free parameter $\alpha$. The hydrostatic equilibrium equations in the EMSG framework are derived and solved numerically to obtain mass-radius relations for soft, stiff, and intermediate equations of state (EOS). Observational measurements of NS masses and radii are used to constrain the fundamental-mode ($f$-mode) oscillation frequency through its universal relation with the tidal Love number and compactness. Results indicate that the Stiff EOS undergoes a phase transition at the highest energy densities and pressures, followed by the Intermediate and Soft EOSs, highlighting the distinctive characteristics of these models. Additionally, the study explores the impact of EOS choice on the sound speed profile of NSs, reaffirming the physical validity of the models across varying $\alpha$ values.}




\maketitle

\section{Introduction}\label{sec1}

\label{intro}
Neutron stars (NSs), the remnants of collapsed massive stars, offer unique opportunities as astrophysical laboratories due to their extreme density and distinctive composition. The intense gravitational forces within their interiors give rise to exotic states of matter, potentially involving quarks and other particles, providing valuable insights into fundamental physics. NSs act as sources of gravitational waves (GWs) \cite{doi:10.1126/science.aap9811,Abbott_2020,PhysRevX.9.011001}, creating ripples in spacetime through the acceleration of massive objects. During various processes, such as merging with other NSs or black holes (BHs) \cite{PhysRevX.9.011001,Fragione_2021}, NSs emit GWs. Detecting these waves offers a unique perspective into the dynamics of NSs and their internal structure. NS mergers, detected through both GWs and electromagnetic signals like gamma-ray bursts \cite{sgrb1,sgrb2}, enable comprehensive studies, laying the foundation for multimessenger astronomy \cite{Abbott_2017}. The distinct imprints on emitted GWs, influenced by different equations of states (EOSs) and compositions within NSs, provide a means to explore the nature of matter under extreme conditions.
\\
In the era of GW astronomy, some particularly important relationships involve the tidal deformability parameters, or simply, tidal deformabilities, of NSs, which are associated with the tidal Love numbers. During a BNS inspiral, the gravitational field of each star induces deformation on the other star through tidal forces. These deformations, described by the tidal deformability parameters, modify the trajectory of each star, leaving an imprint on the resulting GW signal. Thanks to the GW170817 event, the first detection of GWs from a BNS merger has been initiated to elucidate the nature of dense matter at supranuclear densities inside NSs, as well as to place constraints on the canonical NS tidal deformability \cite{GW170817}.
\\
The study of various theories of gravity in the strong-field regime remains a relatively unexplored area \cite{CAPOZZIELLO2011167,OLMO20201}. While Einstein's General Theory of Relativity (GR) has proven to be remarkably effective in describing gravitational interactions across different scales—bolstered by the groundbreaking detection of GWs—there are compelling reasons to investigate alternative theories of gravity. These motivations stem from unresolved issues such as the nature of dark matter and dark energy on galactic and cosmological scales, as well as the presence of singularities in the early Universe and within black holes \cite{PhysRevD.98.024031}.
\\
In the absence of a fundamental quantum gravity theory for describing the complete gravitational action, modified theories of gravity initially focused on simple modifications to the gravitational Lagrangian. One of the earliest approaches involved replacing the linear function of spacetime curvature (the Ricci scalar \(R\)) with an analytic function \(f(R)\). This modification alters the left-hand side of Einstein’s field equations, typically written as \(R_{\mu\nu} - g_{\mu\nu}R/2 = \kappa T_{\mu\nu}\) in standard notation \cite{CAPOZZIELLO2011167}.
Subsequently, the generalized \(f(R, T)\) gravity was proposed \cite{PhysRevD.84.024020}, incorporating not only \(R\) but also the trace \(T\) of the energy-momentum tensor \(T_{\mu\nu}\) of matter. More recently, a more comprehensive covariant formulation has been developed, which considers a nonlinear Lorentz scalar involving the full matter Lagrangian. This theory, known as energy-momentum powered gravity (EMPG), is expressed as \(f(R, T^2) \equiv R + \alpha (T^2)^n = R + \alpha (T_{\mu\nu} T^{\mu\nu})^n\), where \(\alpha\) is a parametric constant \cite{EMSG_OAkarsu, Nari, Akarsu, Katırcı2014, Nazari, Nazari1,Fazlollahi2023}.
The case with the dimensionless parameter \(n = 1\), referred to as energy-momentum squared gravity (EMSG), becomes particularly effective at high-energy densities, such as those encountered in the interiors of NSs \cite{EMSG_OAkarsu,Nari}. EMSG also addresses the limitations of \(f(R, T)\) gravity in scenarios involving a perfect fluid with the equation of state \(P = \rho/3\), where the trace \(T = 0\) reduces \(f(R, T)\) to the standard \(f(R)\) theory. Additionally, EMSG introduces deviations from General Relativity (GR) in low-curvature regimes \cite{Danarianto2023}, while remaining consistent with GR in vacuum.
The field equations in EMSG modify the Tolman-Oppenheimer-Volkoff (TOV) equations, which, in turn, influence the astrophysical properties of NSs.
\\
Recently, the parameter \(\alpha\) in the EMSG model has been constrained \cite{EMSG_OAkarsu} using the 2\(M_\odot\) maximum mass NS constraint and the physicality of certain effective EOS at the centre of NSs, to lie within the range \(-10^{-38} \lesssim \alpha \lesssim 10^{-37} \, \mathrm{cm^3/erg}\). Meanwhile, binary pulsar observations \cite{Akarsu} provide a compatible range of \(-6 \times 10^{-38} \lesssim \alpha \lesssim 10^{-36} \, \mathrm{cm^3/erg}\), whereas constraints from Solar System tests yield significantly smaller values of \(-4 \times 10^{-27} < \alpha < 10^{-26} \, \mathrm{cm^3/erg}\).
\\
NSs exhibit radial \cite{Chandrasekhar_1964, Chanmugam_1977, Kokkotas_2001} and non-radial \cite{mcdermott_1988, Kunjipurayil_2022} oscillations, emitting gravitational waves (GWs) through quasi-normal modes (QNMs). QNMs include $f$-mode \cite{Zhao_2022}, $p$-mode \cite{Sotani_2021}, $g$-mode \cite{Finn_1987}, $r$-mode \cite{Haskell_2014}, and $w$-mode \cite{Benhar_1999}. The frequencies of these oscillations are closely tied to the star's internal structure and composition \cite{Tianqi_2022}. Theoretical studies suggest that the $f$-mode is the most likely oscillation mode to be detected first, contributing approximately 10\% of the gravitational radiation for $l=2$ \cite{Shibagaki_2020}.
\\
Oscillation frequencies provide insights into NS types, e.g., $f$ and $g$-modes suggest hybrid stars \cite{Sotani_2011}. GW frequencies 0-1 kHz may indicate hybrid stars, exceeding 7 kHz strange stars \cite{Flores_2014}. Relationships between frequencies and NS properties, such as compactness \cite{Andersson_1998}, are explored through universal relations (URs), offering model-independent insights \cite{Kent_yagi_2013, Mohanty_2024}.
\\
In this study, we compare the predictions of EMSG theory relative to GR to investigate strong-field effects in NSs. The field equations of EMSG, incorporating the standard physical energy-momentum tensor, can be reformulated to resemble the Einstein field equations of GR, but with an effective or modified energy-momentum tensor. This formulation facilitates the calculations of the compactness, tidal deformation, and other properties of NSs \cite{Rahmansyah}. So, after taking the three different kinds of EOS models, Stiff, Intermediate and Soft \cite{HQPT}, we have generated the data by varying the \(\alpha\) parameters. Then, we established the URs between compactness ($C$) and \(f\)- mode oscillations ($C$-$f$)
 and between Tidal Love number and \(f\)- mode oscillations (\(f\)-Love) relations. Finally, using those URs, we have constrained the \(f\)- mode oscillation frequencies.

This paper is organized as follows; In Sec. \ref{sec:TF}, we described the theoretical framework, inside this Sec. \ref{sec:TF}, in SubSec. \ref{sec:EMSG} we discussed about EMSG and in SubSec. \ref{sec:hydrostatics} Hydrostatic equilibrium in EMSG has been presented. Then, in Sec. \ref{NRO}, we have discussed the Non-Radial Oscillation in Cowling approximation. In Sec. \ref{UR}, we presented the $C-f$ (in SubSec. \ref{CF}) and $f$-Love (in SubSec. \ref{fL}) URs followed by the comparison study of our isotropic (($\alpha=0$ (GR)) for studying EMSG) case with the previous works in SubSec. \ref{comp}. We have shown our final results and a brief discussion about the results in Sec. \ref{RD}. Finally, we have concluded our work in Sec. \ref{conclusion}.

Throughout this paper, we adopt mostly positive signatures (-, +, +, +) and utilize a geometrized unit system ($G=c=\hbar=1$).

\section{Theoretical Framework}
\label{sec:TF}
\subsection{Energy-Momentum Squared Gravity (EMSG)}

\label{sec:EMSG} 
A self-contraction term of the Energy-momentum tensor (EMT), $T_{\mu \nu }T^{\mu \nu }$, is added to the Einstein-Hilbert action with a cosmological constant to create the EMSG model. \cite{Akarsu,EMSG_OAkarsu,OLMO20201,EMSG_NAlam}:
\begin{align}
S=\int \left[\frac{1}{2\kappa}\left(\mathcal{R}-2\Lambda\right)+\alpha T_{\mu\nu}T^{\mu\nu}+ \mathcal{L}_{\mathrm{m}}\right]\sqrt{-g}\,\mathrm{d}^4x,
\label{action}
\end{align}
where the scalar curvature is $\mathcal{R}$, the gravitational coupling is $\kappa =8\pi G$, Newton's constant $G$, a cosmological constant $\Lambda $, and the matter Lagrangian density is $\mathcal{L}_{\mathrm{m}}$. The gravitational coupling strength of the EMSG modification is determined by the term $T_{\mu \nu }T^{\mu \nu }$, which has a real constant $\alpha $.
The EMT in terms of matter Lagrangian density ($\mathcal{L}_{\mathrm{m}}$), is defined as \cite{EMSG_OAkarsu,EMSG_NAlam}
\begin{align}  \label{tmunudef}
T_{\mu\nu}=-\frac{2}{\sqrt{-g}}\frac{\delta(\sqrt{-g}\mathcal{L}_{\mathrm{m}})}{\delta g^{\mu\nu}}=g_{\mu\nu}\mathcal{L}_{\mathrm{m}}-2\frac{\partial \mathcal{L}_{\mathrm{m}}}{\partial g^{\mu\nu}},
\end{align}
which is independent of its derivatives and solely dependent on the metric tensor components. We consider the EMT's ideal fluid form, which is given by
\begin{align}  \label{em}
T_{\mu\nu}=(\mathcal{E}+P)u_{\mu}u_{\nu}+P g_{\mu\nu},
\end{align}
where $\mathcal{E} $ is the energy density, $P$ is the thermodynamic pressure and $u_{\mu }$ is the four-velocity satisfying the conditions $u_{\mu }u^{\mu }=-1$, $\nabla _{\nu }u^{\mu }u_{\mu }=0$. Now, by varying the modified action in EMSG given in Eq. \eqref{action} with respect to the inverse metric, we obtain the modified Einstein's field equations in EMSG as:
\begin{align}
G_{\mu\nu}+\Lambda g_{\mu\nu}=\kappa T_{\mu\nu}+\kappa \alpha \left(g_{\mu\nu}T_{\sigma\epsilon}T^{\sigma\epsilon}-2\theta_{\mu\nu}\right),
\label{fieldeq}
\end{align}
where $G_{\mu \nu }=\mathcal{R}_{\mu \nu }-\frac{1}{2}g_{\mu \nu }\mathcal{R}$ is the Einstein tensor. The new tensor $\theta _{\mu \nu }$ is defined as,
\begin{equation}
\begin{aligned} \theta_{\mu\nu}=& T^{\sigma\epsilon}\frac{\delta
T_{\sigma\epsilon}}{\delta g^{\mu\nu}}+T_{\sigma\epsilon}\frac{\delta
T^{\sigma\epsilon}}{\delta g^{\mu\nu}} \\ =&-2\mathcal{L}_{\rm
m}\left(T_{\mu\nu}-\frac{1}{2}g_{\mu\nu}T\right)-TT_{\mu\nu} \\
&+2T_{\mu}^{\gamma}T_{\nu\gamma}-4T^{\sigma\epsilon}\frac{\partial^2
\mathcal{L}_{\rm m}}{\partial g^{\mu\nu} \partial g^{\sigma\epsilon}}.
\label{theta} \end{aligned}
\end{equation}
Here $T = g^{\mu\nu}T_{\mu\nu}$, is the trace of the EMT. We can see the last term in Eq. \eqref{theta} contains the second order derivative of $\mathcal{L}_{\mathrm{m}}$ but as we have seen that the Eq. \eqref{tmunudef} does not contain any second order derivative so, we can neglect that last term in Eq. \eqref{theta}. 
In the present work, we consider $\mathcal{L}_{\mathrm{m}}=P$ \cite{Faraoni,PhysRevD.109.104055}. The covariant divergence of Eq. \eqref{fieldeq} then becomes
\begin{equation}
\nabla^{\mu}T_{\mu\nu}=-\alpha g_{\mu\nu}\nabla^{\mu}
(T_{\sigma\epsilon}T^{\sigma\epsilon})+2\alpha\nabla^{\mu}\theta_{\mu\nu},
\label{nonconservedenergy}
\end{equation}
we can see that the local covariant energy-momentum is not generally conserved except $\alpha =0$.

Substituting Eq. \eqref{em} in Eq. \eqref{theta}, and then using the resultant equation in Eq. \eqref{fieldeq}, we finally get the equation \cite{EMSG_OAkarsu,EMSG_NAlam}
\begin{align}
G_{\mu\nu}+\Lambda g_{\mu\nu}=\kappa \mathcal{E} \left[\left(1+\frac{P}{\mathcal{E}}\right)u_{\mu}u_{\nu}+\frac{P}{\mathcal{E}}g_{\mu\nu}\right]\\
+\alpha\kappa\mathcal{E}^2\left[2\left(1+\frac{4P}{\mathcal{E}}+\frac{3P^2}{\mathcal{E}^2}\right)u_{\mu}u_{\nu}+\left(1+\frac{3P^2}{\mathcal{E}^2}\right)g_{\mu\nu}\right].
\label{fieldeq2}
\end{align}
The GR Einstein field equation can be obtained by redefining Eq.(\ref{fieldeq2}) as,
\begin{equation}
    G^{\mu\nu} + \Lambda g^{\mu\nu} = \kappa T^{\mu\nu}_{\mathrm{eff}},
\end{equation}
with an effective energy-momentum tensor
\begin{equation}
\label{teff}
    T^{\mu\nu}_{\mathrm{eff}} = (\mathcal{E}_{\mathrm{eff}}+P_{\mathrm{eff}})u^{\mu}u^{\nu} + P_{\mathrm{eff}}g^{\mu\nu}
\end{equation}
for an ideal fluid, where the $\mathcal{E}_{\mathrm{eff}}$ is the effective energy density defined as,
\begin{equation}
\label{Eeff}
    \mathcal{E}_{\mathrm{eff}} = \mathcal{E} + \alpha\mathcal{E}^2\Bigg(1+\frac{8P}{\mathcal{E}} + \frac{3P^2}{\mathcal{E}^2}\Bigg)
\end{equation}
and $P_{\mathrm{eff}}$ is the effective pressure defined as,
\begin{equation}
\label{Peff}
    P_{\mathrm{eff}} = P + \alpha\mathcal{E}^2\Bigg(1+\frac{3P^2}{\mathcal{E}^2}\Bigg)
\end{equation}

From the study of \cite{EMSG_OAkarsu}, we get that they have claimed, $|\alpha |\sim\mathcal{E} ^{-1}$ and hence the corrections due to the EMSG modifications would be observable for compact objects like NSs if $|\alpha |\sim 10^{-37}\,\mathrm{erg^{-1}\,cm^{3}}$. So, we can use NSs as a probe to make constraints on $\alpha $.

\subsection{Hydrostatic equilibrium in EMSG}

\label{sec:hydrostatics}
The line element that describes the space-time inside a spherically symmetric  star is \cite{Wald:1984rg,Schwarzschild:1916uq}
\begin{equation}  
\label{eqn:metric}
\mathrm{d} s^2 = -e^{2\nu\left(r\right)}\mathrm{d} t^2 +e^{2\lambda\left(r\right)}\mathrm{d} r^2+r^2\mathrm{d}\theta^2+r^2\sin^2\theta \, \mathrm{d}\phi^2
\end{equation}
wheres $\nu (r)$ and $\lambda ( r) $ are the metric functions. Using the metric provided in Eq.\eqref{eqn:metric} and substituting it into Eq.\eqref{fieldeq2}, we arrive at the following set of field equations, given in Eqs.~\eqref{eqn:f1}-\eqref{eqn:f2}:

\begin{equation}
\begin{aligned}
\frac{1}{r^2} - \frac{e^{-2\lambda}}{r^2} \left(1 - 2r \frac{\mathrm{d} \lambda}{\mathrm{d} r} \right) = \kappa \mathcal{E} + \kappa \alpha \mathcal{E}^2 \left(1 + 8 \frac{P}{\mathcal{E}} + 3 \frac{P^2}{\mathcal{E}^2} \right),
\end{aligned}
\label{eqn:f1}
\end{equation}
\begin{equation}
\begin{aligned}
-\frac{1}{r^2} + \frac{e^{-2\lambda}}{r^2} \left(1 + 2r \frac{\mathrm{d} \nu}{\mathrm{d} r} \right) = \kappa P + \kappa \alpha \mathcal{E}^2 \left(1 + 3 \frac{P^2}{\mathcal{E}^2} \right),
\end{aligned}
\label{eqn:f2}
\end{equation}

where $\mathcal{E} $ and $P$ are the energy density and pressure at the distance $r$ from the center of NS.

Now by solving the modified Einstein equations in EMSG, we will get the modified Tolman-Oppenheimer-Volkoff (TOV) equations as,
\begin{align}
\frac{\mathrm{d} m}{\mathrm{d} r}=4\pi r^2 \mathcal{E} \left[1+\alpha\mathcal{E} \left(1+8\frac{P}{\mathcal{E}}+3\frac{P^2}{\mathcal{E}^2 }\right)\right],  \label{TOV1}
\end{align}
\begin{align}
\frac{\mathrm{d} P}{\mathrm{d} r}&= -\frac{m\mathcal{E} }{r^2}\left(1+\frac{P}{\mathcal{E}}\right) \left( 1-\frac{2m}{r}\right)^{-1}  \notag \\
&\times \left[ 1+\frac{4\pi r^3 P}{m }+\alpha \frac{4\pi r^3\mathcal{E}^2}{m}\left(1+3\frac{P^2}{\mathcal{E}^2}\right)\right] \notag \\
&\times \left[1+2\alpha\mathcal{E}\left(1+3\frac{P}{\mathcal{E}} \right)\right] \left[1 + 2\alpha\mathcal{E} \left(c_s^{-2}+3\frac{P}{\mathcal{E}}\right)\right]^{-1},  \label{TOV2}
\end{align}
\begin{align}  \label{TOV3}
\frac{\mathrm{d} \nu}{\mathrm{d} r} =&- \left\{\mathcal{E} \left( 1+\frac{P}{\mathcal{E}}\right) \left[1+2\alpha\mathcal{E}\left( 1+3\frac{P}{\mathcal{E}} \right)\right] \right\}^{-1}  \notag \\
&\times \left[ \left(1+6\alpha P\right) \frac{\mathrm{d} P}{\mathrm{d} r}+2\alpha\mathcal{E}\frac{\mathrm{d}\mathcal{E}}{\mathrm{d} r} \right],
\end{align}
\begin{figure*}[h!]
    \includegraphics[width=0.5\linewidth]{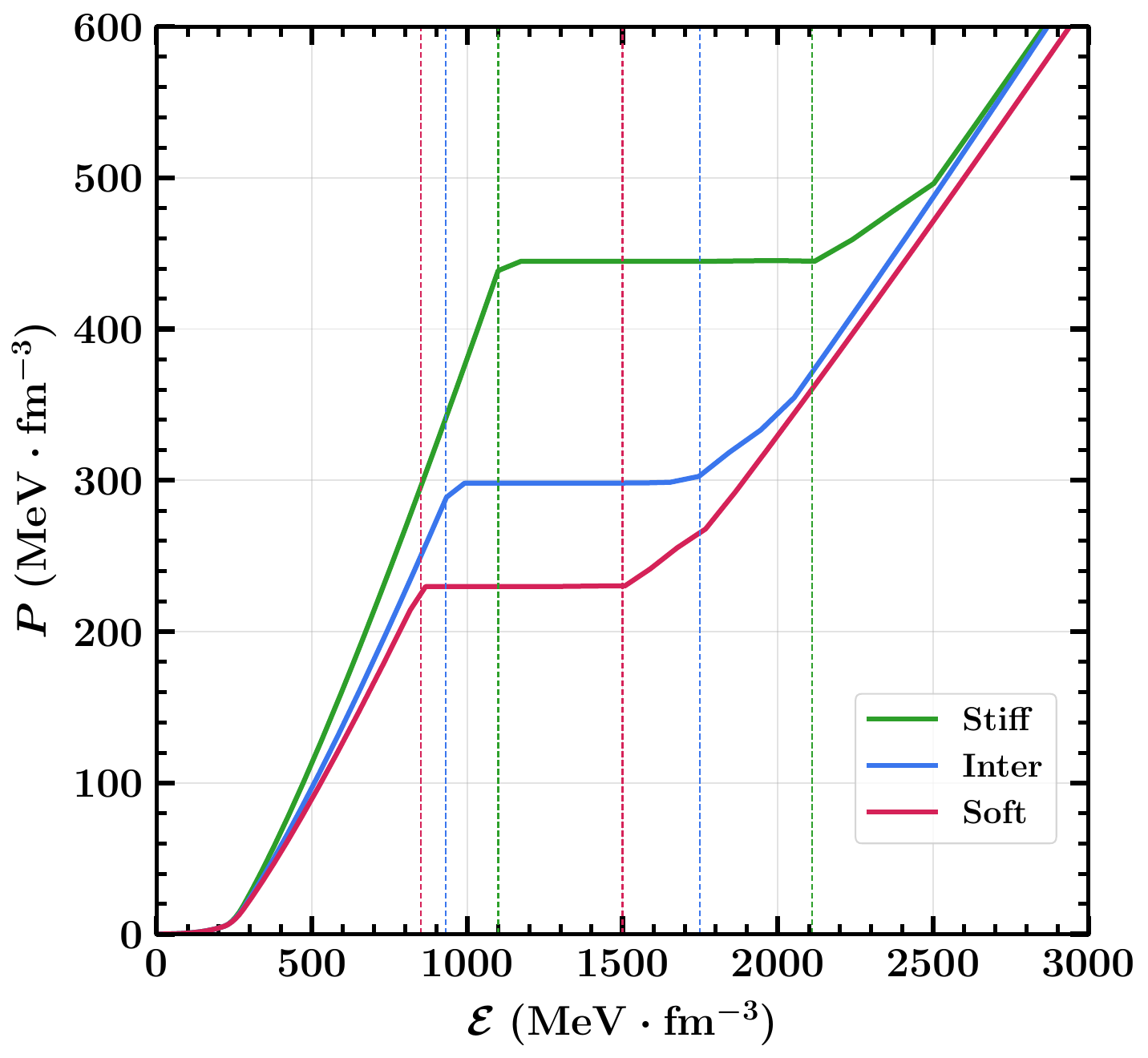}
    \includegraphics[width=0.5\linewidth]{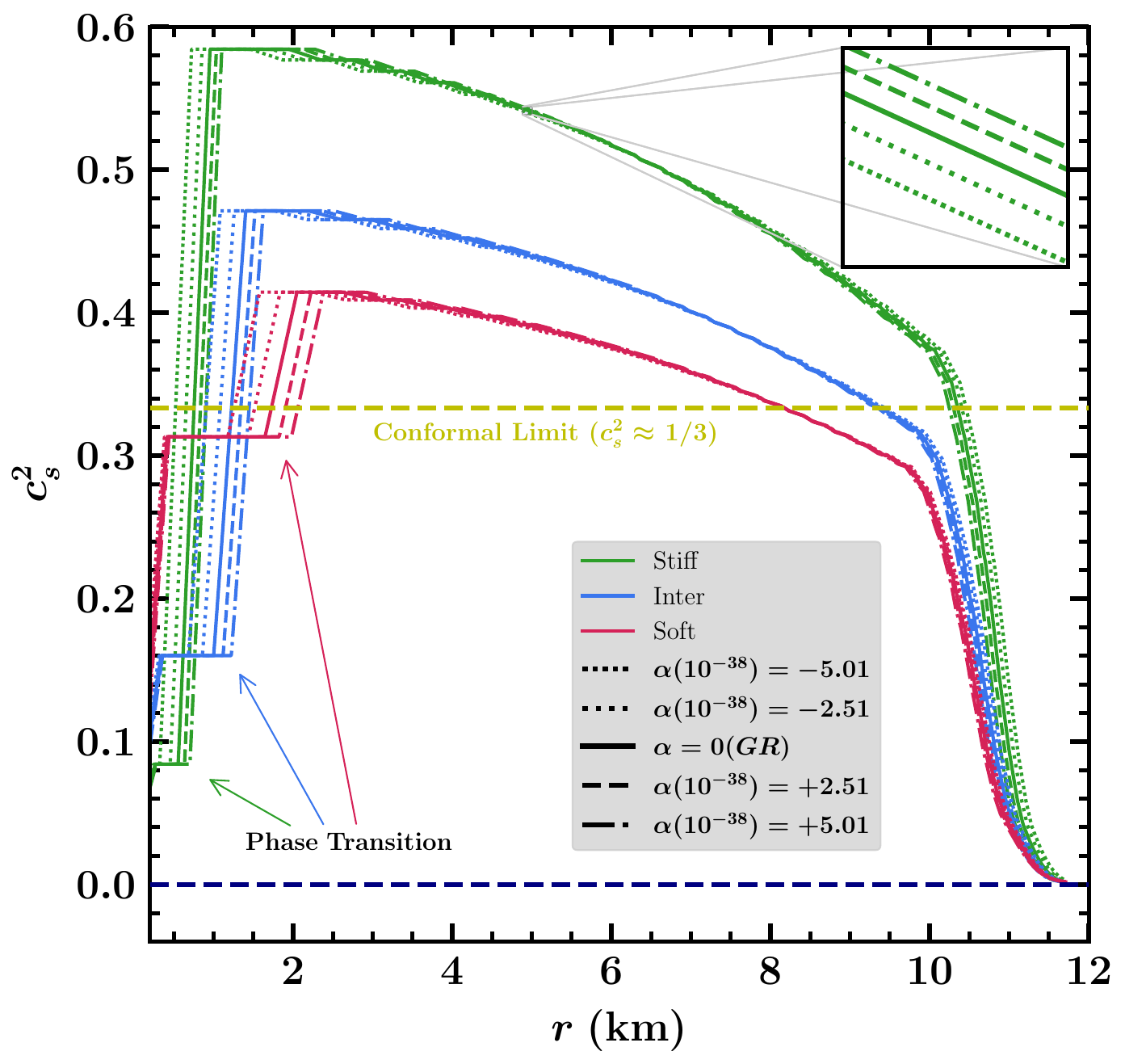}
    \caption{\textit{Left:}Variation of pressure ($P$) with energy density ($\mathcal{E}$) for stiff, intermediate and soft EOSs, represented by different colours. Vertical dashed lines correspond to the phase transition regions for different EOSs. \textit{Right:}The radial profile of sound speed ($c^2_s$) for alpha ($\alpha$) values of the maximum mass NS corresponding to respective EOSs. Zoom plot refers to the clear visibility of the effect of $\alpha$.}
    \label{EOS}
\end{figure*}
 This set of equations, Eqs.~\eqref{TOV1}-\eqref{TOV2}, is closed by an EOS, $P(\mathcal{E})$, which prescribes the relation between the pressure $P(r)$ and the density $\mathcal{E}(r)$. The variation of $P$ as a function of 
$\mathcal{E}$ is represented in the Left panel of Fig. \ref{EOS}. $c_{s}^2 \equiv \frac{\mathrm{d} P}{\mathrm{d} \mathcal{E}}$ is the sound speed, presented in the Right panel of Fig. \ref{EOS}. $m(r)$ is the enclosed mass corresponding to radius $r$ and $\lambda(r)$ is the metric function defined as
\begin{equation*}
    e^{-2 \lambda}=1-\frac{2 m}{r}.
\end{equation*}
In order to numerically solve the aforementioned set of coupled ordinary differential equations (ODEs), specific boundary conditions need to be established. Conventionally, the surface of the star is set at $r=R$, where the radial pressure becomes zero ($P=0$). As the equilibrium system exhibits spherical symmetry, the Schwarzschild metric is employed to describe the exterior space-time. This choice ensures metric continuity at the surface of the NS and imposes a boundary condition on $\nu$. Specifically, the value of $\nu$ at $r=R$ must coincide with the value of $\nu$ in the Schwarzschild metric at $r=R$.
\begin{equation*}
\nu \left(r=R\right)=\frac{1}{2} \ln \left[1-\frac{2  M}{ R}\right] \, .
\end{equation*}
By making a selection of an EOS governing the radial pressure ($P$) and adopting the EMSG model for $\alpha$, it becomes feasible to numerically solve Eq. (Eqs.~\eqref{TOV1}-\eqref{TOV2}). This numerical solution involves specifying a central energy density $\mathcal{E}(r=0)=\mathcal{E}_c$, while enforcing the initial condition $m(r=0)=0$.
\section{Non-Radial Oscillation in Cowling approximation}
\label{NRO}
GWs, including the fundamental $f$-mode, are emitted when NSs oscillate due to external or internal perturbations \citep{PhysRevD.66.104002}. Nonradial oscillations of spherically symmetric NSs can be studied using the Cowling approximation \citep{Cowling_1941}, which assumes a static spacetime by neglecting metric perturbations. These oscillations, characterized by real eigenvalues, are purely harmonic and provide valuable insights into the internal structure and dynamics of NSs.
\\
The Lagrangian displacement vector of the fluid is given by \\
\begin{equation}
\xi^{i}=\frac{1}{r^2}\Big(e^{-\lambda (r)}W (r),-V
(r)\partial_{\theta},
     -\frac{V(r)}{ \sin^{2}{\theta}}\  \partial _{\phi}\Big)
e^{i\omega t}Y_{lm}(\theta,\phi) ,
\end{equation}
 where $Y_{lm}(\theta,\phi)$ represents the spherical harmonics. The Lagrangian, defined by the functions $V(r)$ and $W(r)$ and governed by the frequency $\omega$, is determined through the solution of the following system of ordinary differential equations\citep{PhysRevD.83.024014},
\begin{eqnarray}
\frac{d W(r)}{dr}&=&\frac{d {\cal E}_{\mathrm{eff}}}{dP_{\mathrm{eff}}}\left[\omega^2r^2e^{\lambda
(r)-2\nu (r)}V(r)
+\frac{d \nu(r)}{dr} W (r)\right] \nonumber \\
&&
-l(l+1)e^{\lambda (r)}V (r) \nonumber \\
\frac{d V(r)}{dr} &=& 2\frac{d\nu (r)}{dr} V
(r)-\frac{1}{r^2}e^{\lambda (r)}W (r) ,
\label{eqn:cowling}
\end{eqnarray}
where, $\cal {E}_{\mathrm{eff}}$ and $P_{\mathrm{eff}}$ are given by Eqs. (\ref{Eeff}) and (\ref{Peff}). So by replacing $\cal {E}_{\mathrm{eff}}$ and $P_{\mathrm{eff}}$ we will get ,
\begin{eqnarray}
\frac{d W(r)}{dr}&=&\frac{d \Bigg[\mathcal{E} + \alpha\mathcal{E}^2\Bigg(1+\frac{8P}{\mathcal{E}} + \frac{3P^2}{\mathcal{E}^2}\Bigg)\Bigg]}{d \Bigg[P + \alpha\mathcal{E}^2\Bigg(1+\frac{3P^2}{\mathcal{E}^2}\Bigg)\Bigg]}\Bigg[\omega^2r^2e^{\lambda
(r)-2\nu (r)}V(r)  \nonumber \\
&&+\frac{d \nu(r)}{dr} W (r)\Bigg] 
-l(l+1)e^{\lambda (r)}V (r) \nonumber \\
\frac{d V(r)}{dr} &=& 2\frac{d\nu (r)}{dr} V
(r)-\frac{1}{r^2}e^{\lambda (r)}W (r) ,
\label{eqn:cowling1}
\end{eqnarray}
where $\nu(r)$ \& $\lambda(r)$ represent metric functions, with the fixed background metric
\begin{equation}
\label{eq:metric}
ds^2=-e^{2 \nu}d t^2+e^{2 \lambda} d r^2+r^2 (d \theta^2+\sin ^2 \theta d \phi^2).
\end{equation}
\begin{figure*}[h!]
    \includegraphics[width=0.5\linewidth]{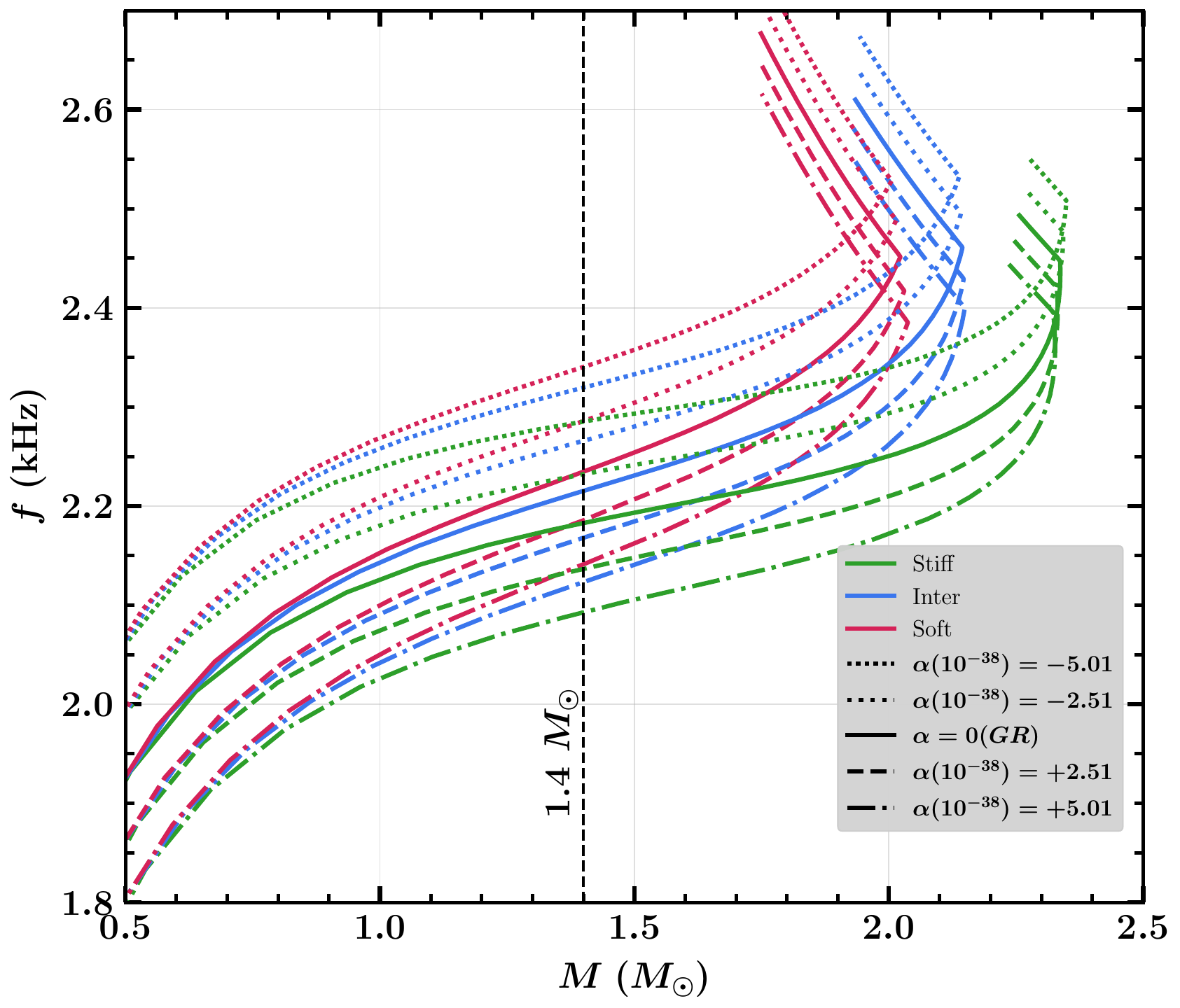}
    \includegraphics[width=0.5\linewidth]{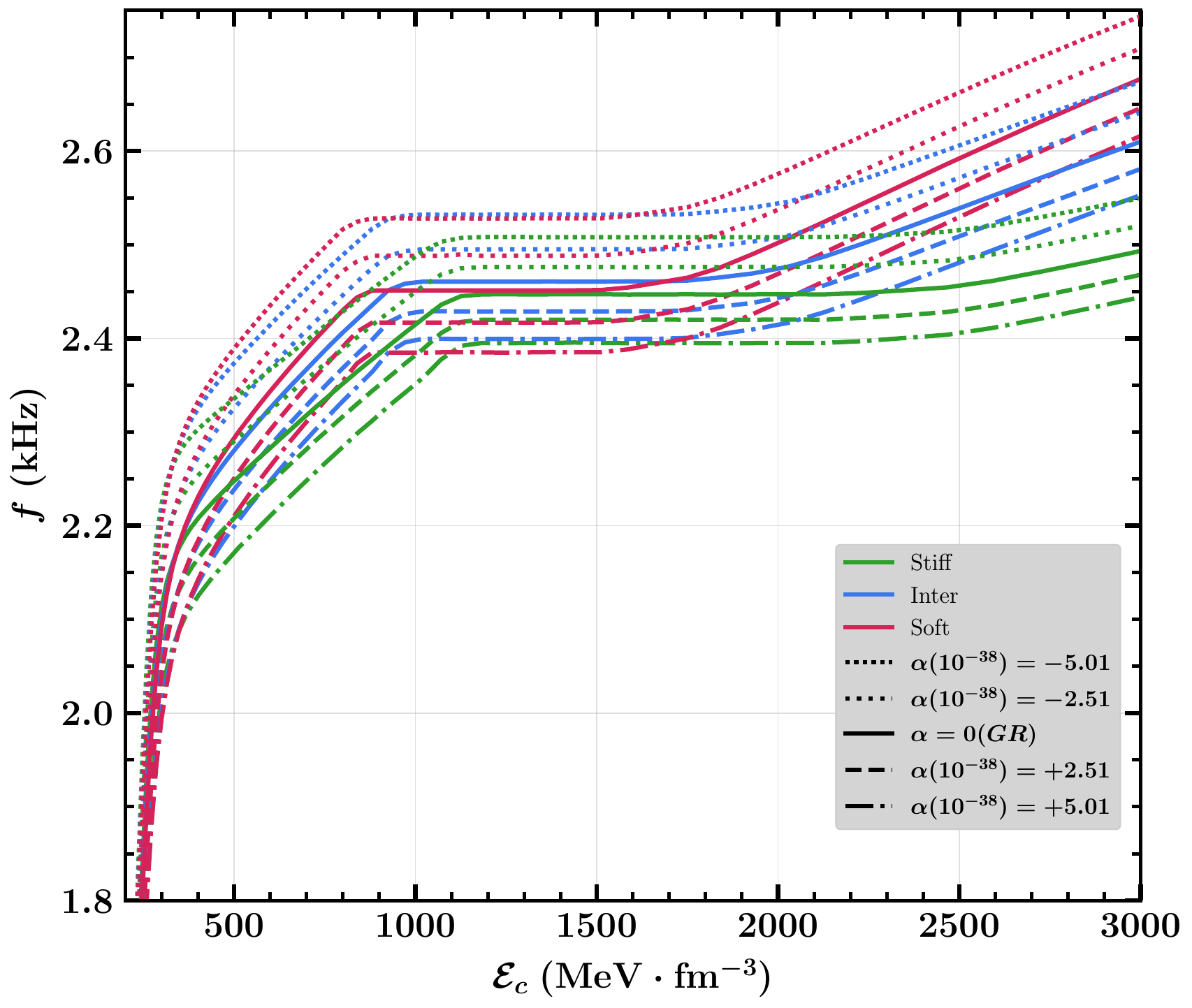}
    \caption{\textit{Left:}Variation of $f$-mode frequency with the mass of the NSs considering the variation of $\alpha$ parameter for different EOSs.\textit{Right:} Variation of $f$-mode frequency with the central energy density.
    }
    \label{f_mode}
\end{figure*}
In the close vicinity of the origin, the solution to Eq. (\ref{eqn:cowling}) exhibits the following behaviour:
\begin{equation}
     W (r)=Br^{l+1}, \ V (r)=-\frac{B}{l} r^l,
\label{eq:bc1}
\end{equation}
where $B$ is an arbitrary constant. To ensure that the perturbation pressure becomes zero at the outer boundary of the star's surface, we need to apply the following additional boundary condition,
\begin{equation}
     \omega^2 e^{\lambda (R)-2\nu (R)}V (R)+\frac{1}{R^2}\frac{d\nu
(r)}{dr}\Big|_{r=R}W (R)=0.
\label{eq:bc2}
\end{equation}
Using the boundary conditions in Eqs. (\ref{eq:bc1}) and (\ref{eq:bc2}), we solve Eq.(\ref{eqn:cowling1}) to obtain the eigenfrequencies of NSs, focusing on the quadrupolar modes ($l=2$). Fig. \ref{f_mode} shows the $f$-mode frequency as a function of NS mass and central energy density for varying $\alpha$ parameters with Stiff, Intermediate, and Soft EOSs. For a given mass, the frequency decreases with positive $\alpha$ and increases with negative $\alpha$ until the maximum stable mass is reached. Utilizing the tidal deformability limit from GW170817 and GW190814, we can constraint the canonical $f$-mode frequency for different $\alpha$ values, as discussed in SubSec. \ref{fL}.

\section{Universal Relations (URs)}
\label{UR}
The primary objective of URs is to investigate the properties of NSs that are challenging to measure through observational means. Numerous URs have been proposed to estimate the properties of NS, with most studies concentrating on isotropic cases \cite{PhysRevD.101.124006, Breu_2016, PhysRevD.91.044034, Yagi:2013bca, Kent_yagi_2013, staykov2016}. In our recent research \cite{Mohanty_2024}, we examined the URs for anisotropic NSs. While some studies have analyzed NSs within the framework of EMSG \cite{EMSG_NAlam, EMSG_OAkarsu}, the investigation of URs for NSs in the context of EMSG remains unexplored. Consequently, this study aims to explore various types of URs on the tidal deformability, compactness, and $f$-mode frequency of NSs by varying the $\alpha$ parameter.

\subsection{$f$-Love relation}
\label{fL}
\begin{figure*}
    \includegraphics[width=0.329\linewidth]{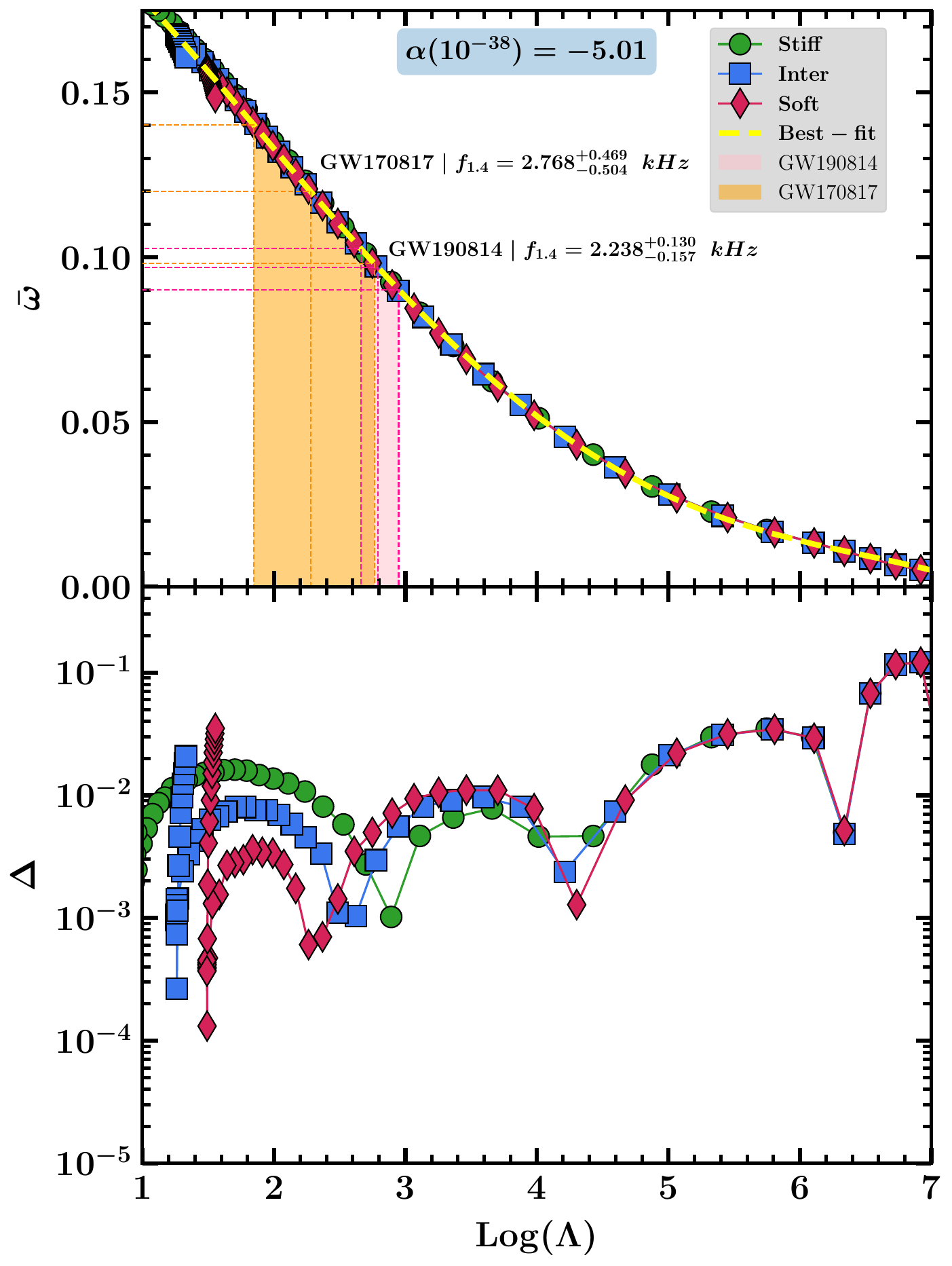}
    \includegraphics[width=0.329\linewidth]{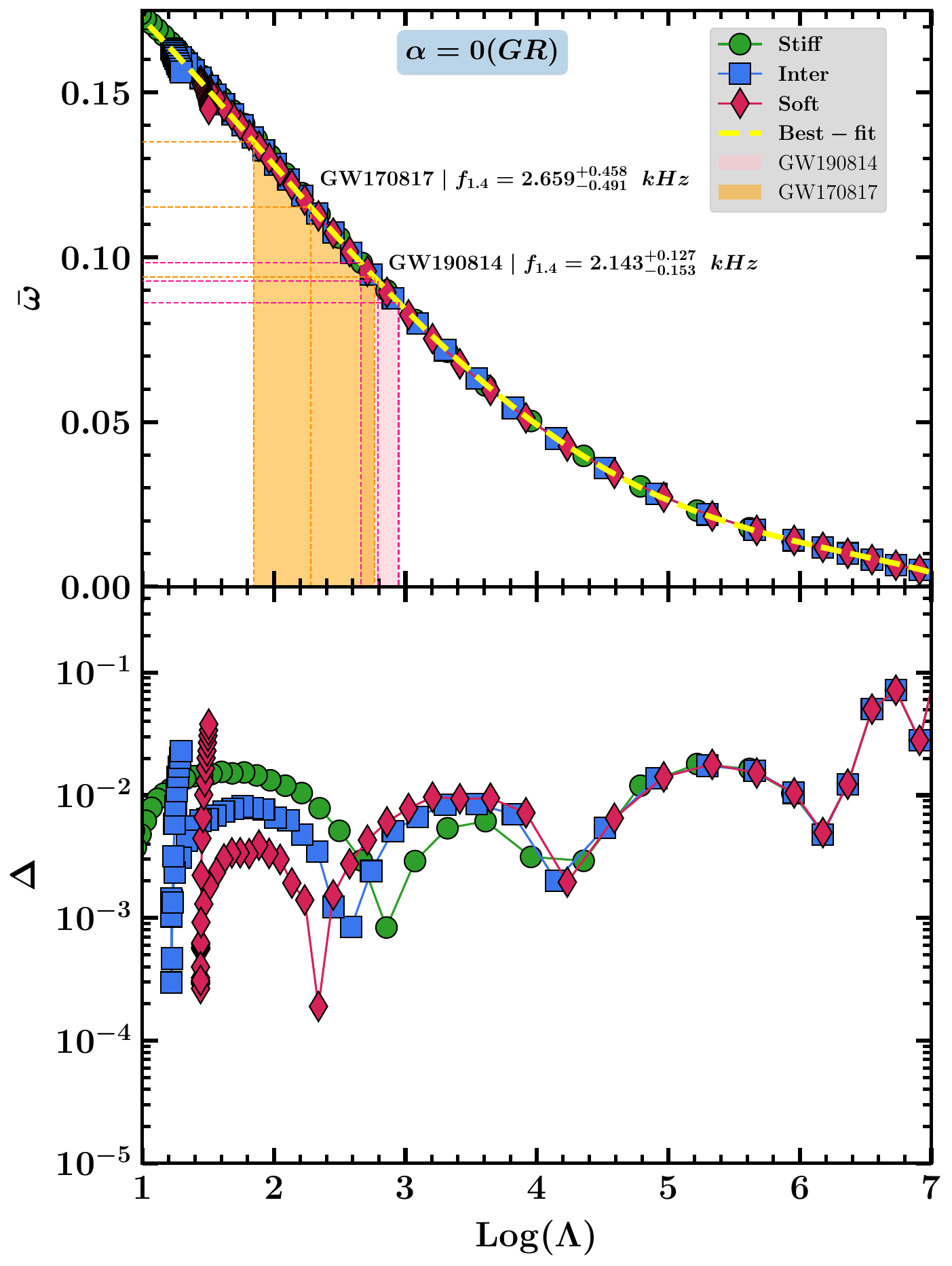}
    \includegraphics[width=0.329\linewidth]{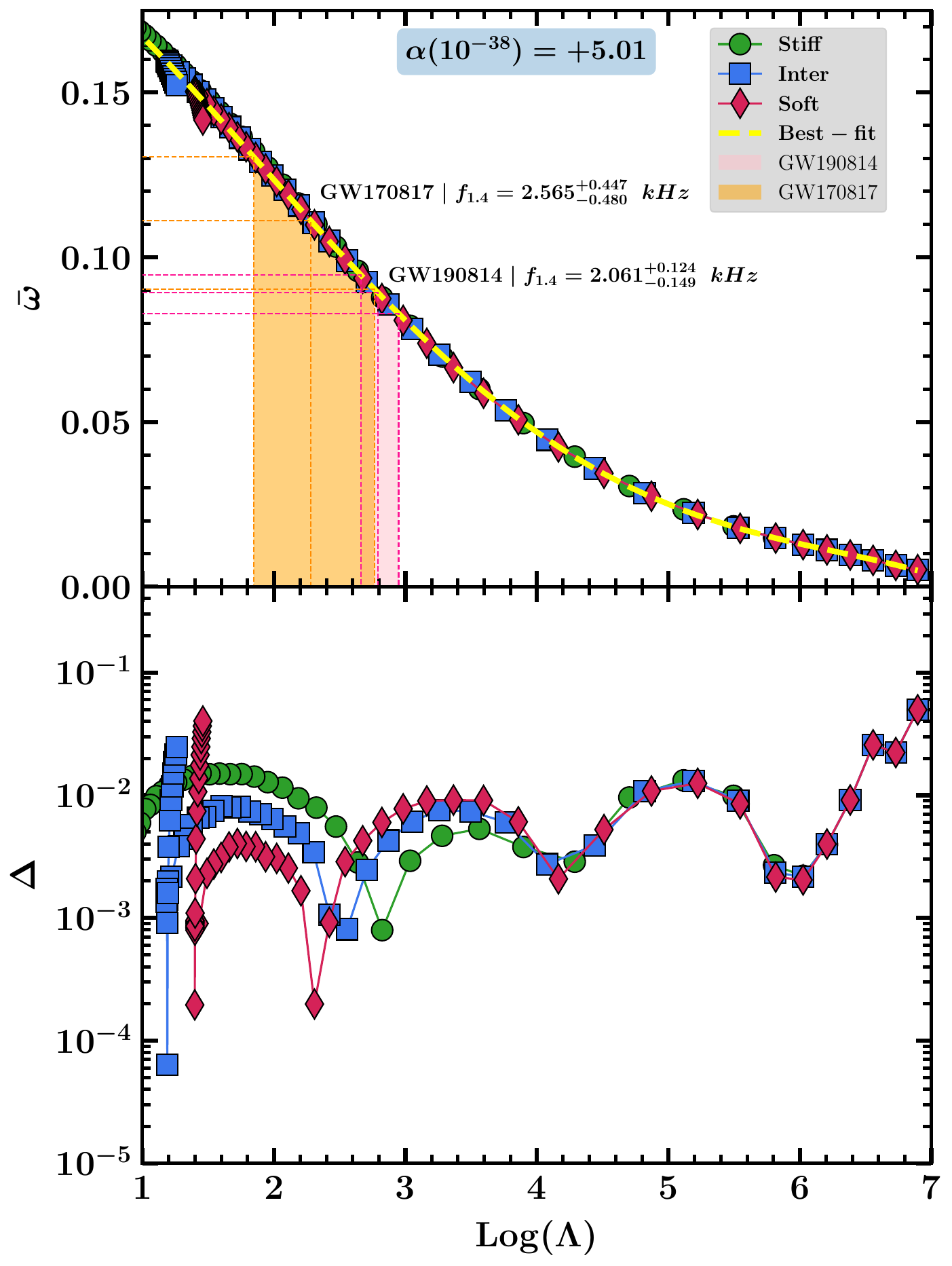}

    \caption{$f$-Love relation with the variation of $\alpha$ parameter for three EOSs. Here the figures are for $\alpha(10^{-38}) =$ $-5.01$, $0$ \& $+5.01$. The yellow dashed line is the best-fit line calculated from Eq.(\ref{eq:f-Love_fitting}). 
    The Right panels show the residuals for the fitting, which are calculated.
    }
    \label{flr1}
\end{figure*}
    

\begin{table*}[b!]
\caption{The fitting coefficients are listed for  $f$-Love relation with $\alpha (10^{-38}) = -5.01, -2.51, 0,  2.51, 5.01$. The reduced chi-squared ($\chi_r^2$) is also given for all cases.}
\centering
\setlength{\tabcolsep}{10pt}
\renewcommand{\arraystretch}{1.5}
\scalebox{0.8}{
    \begin{tabular}{cccccc}
        \hline \hline
        $\alpha (10^{-38}) =$ & $-5.01$ & $-2.51$ & 0  & $+2.51$ & $+5.01$   \\
        \hline
        $d_0\left(10^{-1}\right)=$ & 2.126 & 2.082 & 2.044 & 2.006 & 1.973 \\
        $d_1\left(10^{-2}\right)=$ & -2.272 & -2.121 & -2.012 & -1.898 & -1.827   \\
        $d_2\left(10^{-2}\right)=$ & -1.427 & -1.476 & -1.507 & -1.538 & -1.550  \\
        $d_3\left(10^{-3}\right)=$ & 3.247 & 3.354 & 3.427 & 3.495 & 3.526\\
        $d_4\left(10^{-4}\right)=$ & -1.927 & -2.008 & -2.067 & -2.119 & -2.148 \\
        $\chi_r^2\left(10^{-6}\right)=$ & 1.461 & 1.483 & 1.491 & 1.544 & 1.573  \\
        \hline \hline
    \end{tabular}}
    \label{f-L}
\end{table*}
\begin{table*}[t!]
\centering
\setlength{\tabcolsep}{10pt}
    \caption{The canonical normalized $f$-mode frequency $(\Bar{\omega}_{1.4})$, and f-mode frequency ($f_{1.4}$ in kHz) inferred from GW170817 and GW190814 data.}
    \renewcommand{\arraystretch}{1.8}
    \scalebox{0.8}{
    \begin{tabular}{cccccc}
        \hline \hline & \multicolumn{2}{c}{$\mathrm{GW170817}$} & & \multicolumn{2}{c}{$\mathrm{GW190814}$}  \\
        \hline
         $\alpha (10^{-38})$ & $\bar{\omega}_{1.4}$ & $f_{1.4}$ & & $\bar{\omega}_{1.4}$ & $f_{1.4}$ \\
        \hline 
        $-5.01$ & $0.120^{+0.020} _ {-0.022}$ & $2.768^{+0.469} _ {-0.504}$ & & $0.097^{+0.006} _ {-0.007}$ & $2.238^{+0.130} _ {-0.157}$ \\
    
        $-2.51$ & $0.118^{+0.020} _ {-0.022}$ & $2.712^{+0.463} _ {-0.498}$ & & $0.095^{+0.006} _ {-0.007}$ & $2.189^{+0.128} _ {-0.155}$ \\
        
        0   & $0.115^{+0.020} _ {-0.021}$ & $2.659^{+0.458} _ {-0.491}$ & & $0.093^{+0.005} _ {-0.007}$ & $2.143^{+0.127} _ {-0.153}$ \\
    
        $+2.51$ & $0.113^{+0.020} _ {-0.021}$ & $2.611^{+0.452} _ {-0.485}$ & & $0.091^{+0.005} _ {-0.007}$ & $2.100^{+0.125} _ {-0.151}$ \\
    
        $+5.01$ & $0.111^{+0.019} _ {-0.021}$ & $2.565^{+0.447} _ {-0.480}$ & & $0.089^{+0.005} _ {-0.006}$ & $2.061^{+0.124} _ {-0.149}$ \\
        \hline \hline
    \end{tabular}}
    \label{tab:canonical_f-mode}
    \end{table*}
A critical approach for examining the oscillatory behavior of NSs through observational methods is the establishment of a UR between the non-radial $f$-mode frequency—a potential source of GWs—and tidal deformability, which can be inferred from GW data. The initial investigation into the multi-polar URs between the $f$-mode frequency and tidal deformability of compact stars was conducted by Chan et al. \cite{Chan_2014}, and subsequently refined by Pradhan et al. \cite{Bikram_2023}. Recently, Sotani and Kumar \cite{Sotani_2021} introduced a UR connecting quasi-normal modes and tidal deformability for isotropic ($\alpha=0$ (GR)) NSs. In this study, we determine the $f$-Love relations for NSs in EMSG and utilize a least-squares fit to approximate the relation using the following formula 
\begin{equation}
\label{eq:f-Love_fitting}
    \Bar{\omega} = \sum_{n=0}^{n=4} d_n (\log(\Lambda))^n \, .
\end{equation}
The relationship between $\bar{\omega}$ and $\Lambda$ are presented in Fig. \ref{flr1} for $\alpha (10^{-38}) = -5.01, 0 \& +5.01$.
The coefficients ($d_n$) including $\chi_r^2$ errors are provided in Table \ref{f-L}. For positive values of $\alpha$, the errors in $\chi_r^2$ increase, conversely, errors decrease for negative values of $\alpha$, indicating an EOS-insensitive relation. Consequently, negative $\alpha$ values enhance the $f$-Love UR. Utilizing the tidal deformability constraints of a canonical mass NS from the GW170817 \cite{GW170817} ($\Lambda_{1.4} = 190_{-120}^{+390}$) and GW190814 \cite{GW190814} ($\Lambda_{1.4}= 616_{-158}^{+273}$) events, we establish theoretical bounds on the $f$-mode frequency for each event under varying $\alpha$ parameter, deploying the $f$-Love URs derived in this work. We consider the latter event (GW190814) as a NS-BH merger event, as substantiated by Ref. \cite{Roupas2021} to validate our results. The pink and orange regions in Fig. \ref{flr1}, 
illustrates the tidal constraints for the GW190814 and GW170817 events, respectively, with the corresponding vertical lines depicting the constrained $\bar{\omega}$ values. The canonical $f$-mode frequency determined by the GW170817 and GW190814 events for different $\alpha$ values as obtained in this study is detailed in Table \ref{tab:canonical_f-mode}.
\subsection{$C$-$f$ relation}
\label{CF}
\begin{figure*}
    
    \includegraphics[width=0.329\linewidth]{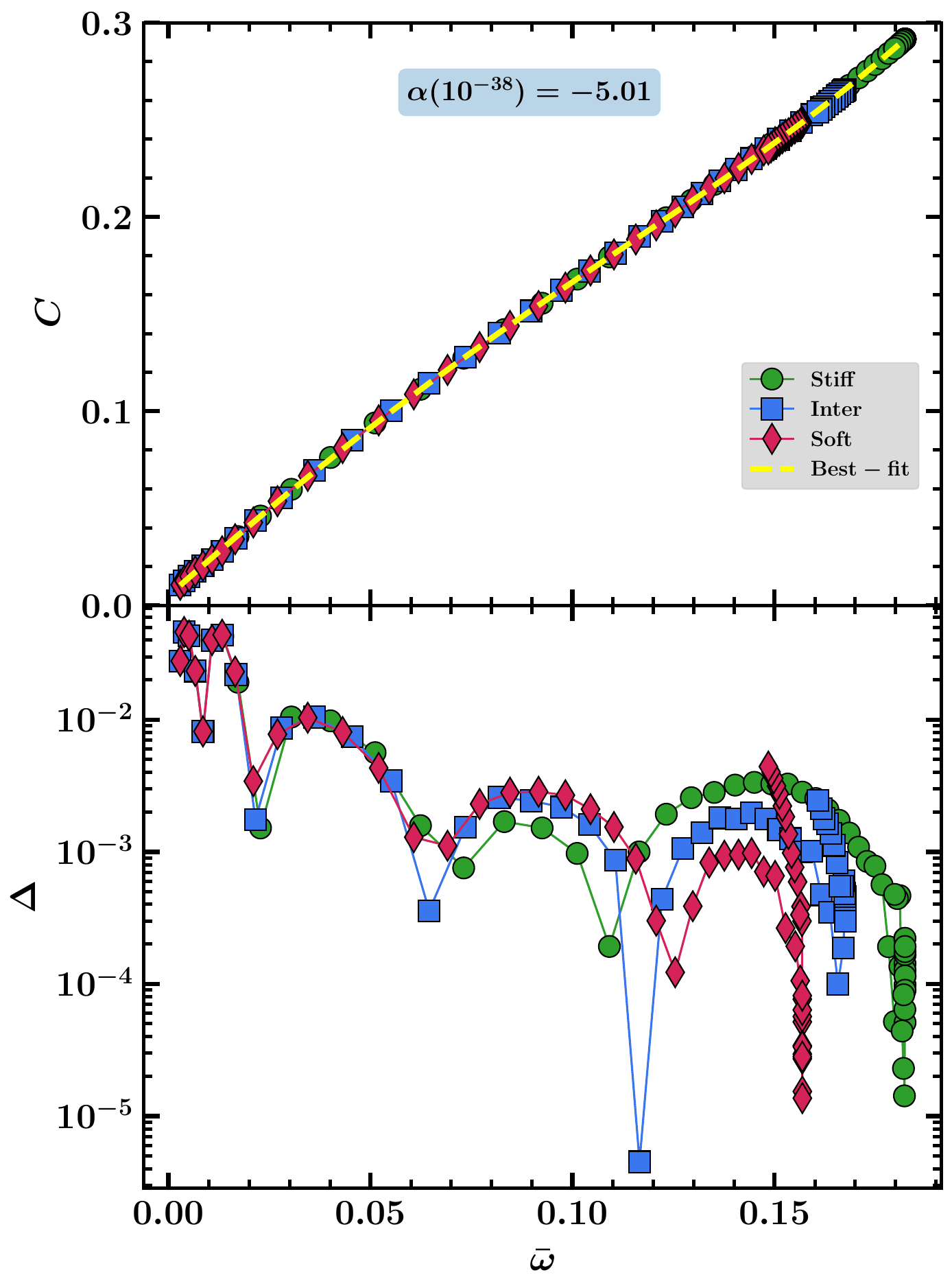}
    \includegraphics[width=0.329\linewidth]{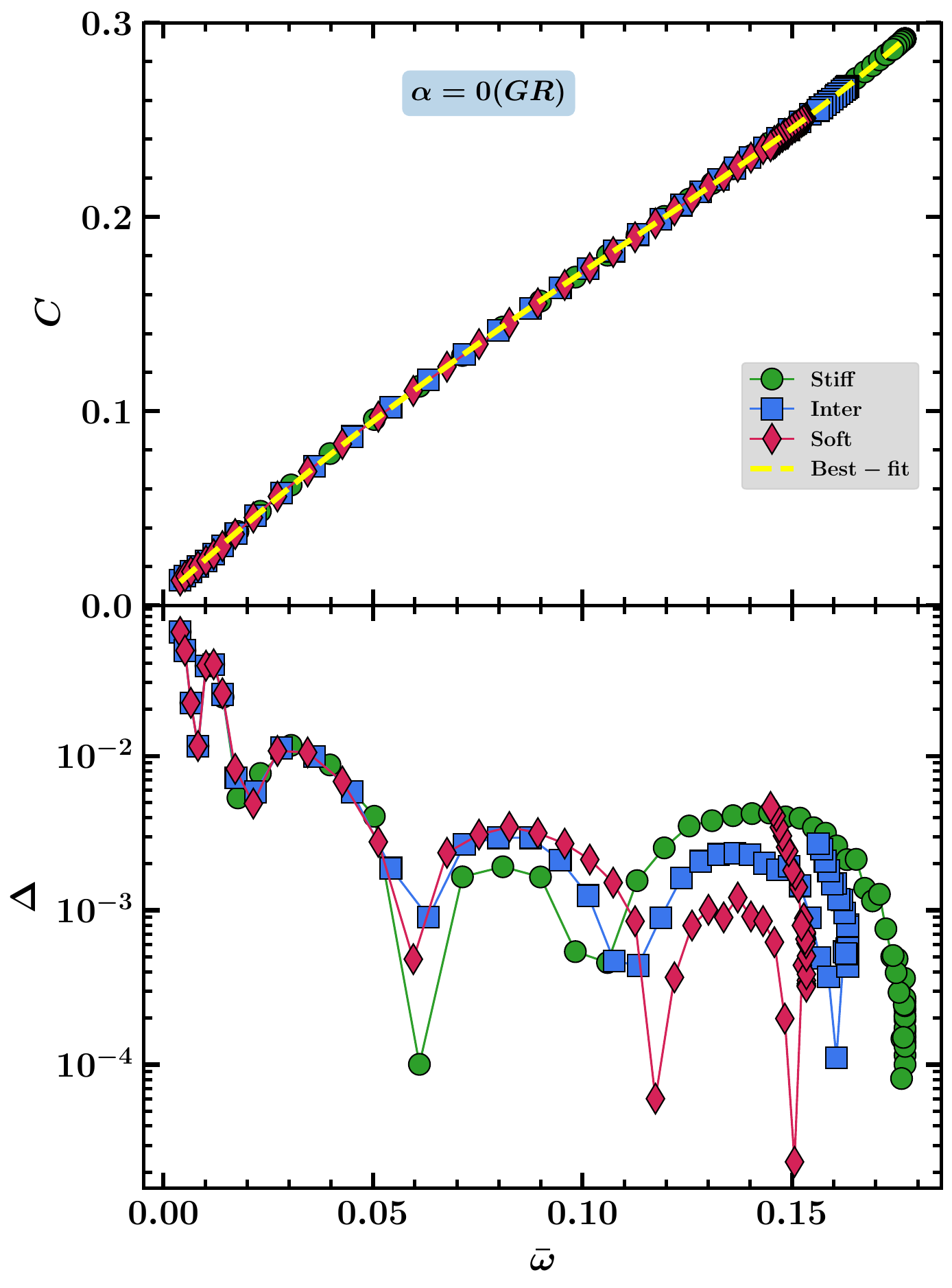}
    \includegraphics[width=0.329\linewidth]{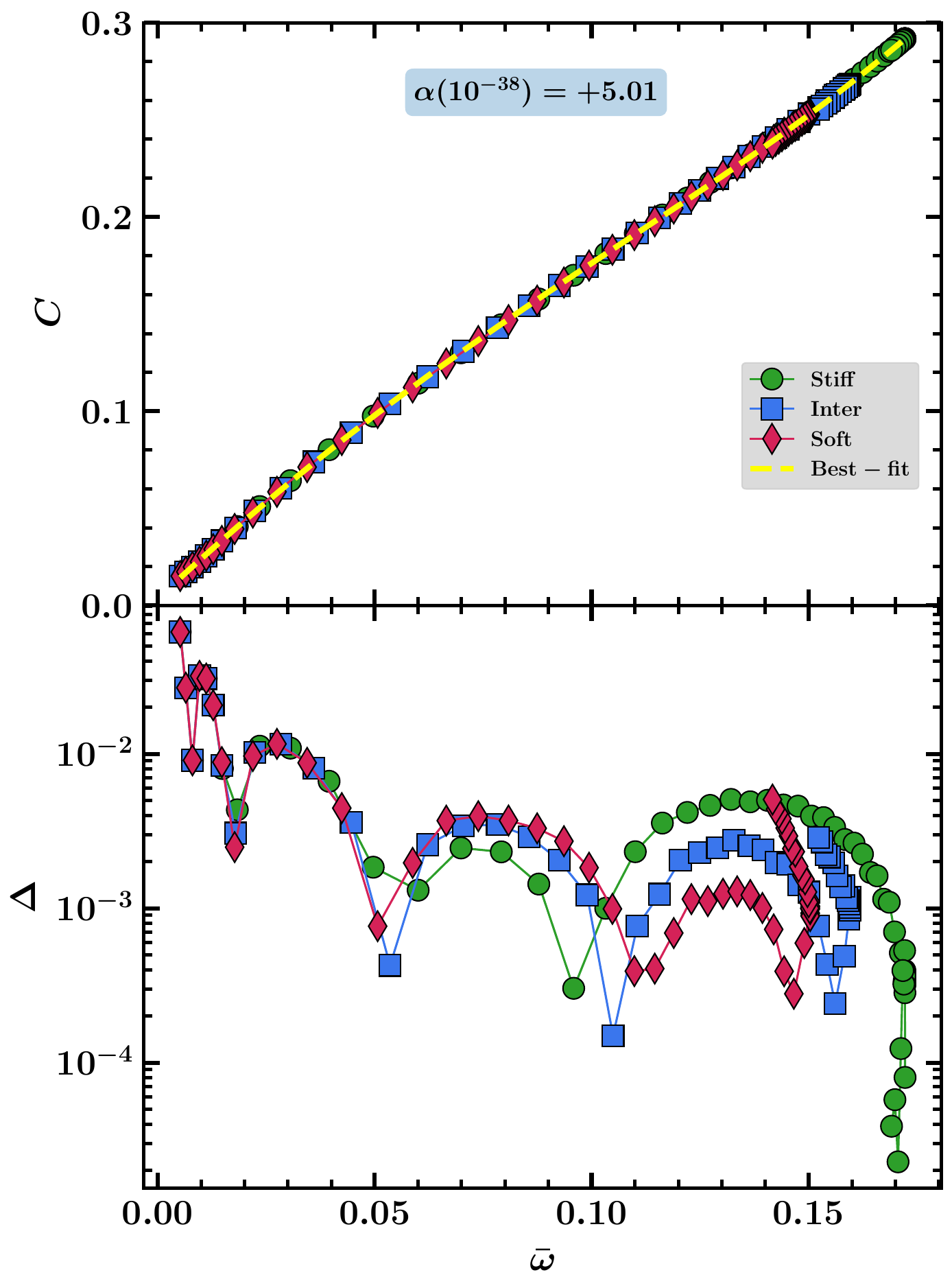}

    \caption{$C$-$f$ relation with the variation of $\alpha$ parameter for three EOSs. Here the figures are for $\alpha(10^{-38}) =$ $-5.01$, $0$ \& $+5.01$. The yellow dashed line is the best-fit line calculated from Eq.(\ref{eq:C-f_fitting}). 
    The Right panels show the residuals for the fitting, which are calculated.
    }
    \label{cf1}
\end{figure*}
    
    
    

\begin{table*}[h!]
\caption{The fitting coefficients are listed for  $C$-$f$ relation with $\alpha (10^{-38}) = -5.01, -2.51, 0, +2.51, +5.01$. The reduced chi-squared ($\chi_r^2$) is also given for all cases.}
\centering
\setlength{\tabcolsep}{6pt}
\renewcommand{\arraystretch}{1.5}
\scalebox{0.8}{
    \begin{tabular}{cccccc}
        \hline \hline
        $\alpha (10^{-38}) =$ & $-5.01$ & $-2.51$ & 0  & $+2.51$ & $+5.01$   \\
        \hline
        $b_0\left(10^{-3}\right)=$ & 4.988 & 4.696 & 4.358 & 3.987 & 3.565 \\
        $b_1=$ & 1.846 & 1.902 & 1.961 & 2.022 & 2.088   \\
        $b_2=$ & -1.842 & -2.270 & -2.769 & -3.355 & -4.103  \\
        $b_3\left(10^{1}\right)=$ & -1.252 & -1.103 & -0.899 & -0.626 & 0.205\\
        $b_4\left(10^{1}\right)=$ & 7.885 & 7.927 & 7.844 & 7.585 & 6.899 \\
        $\chi_r^2\left(10^{-7}\right)=$ & 1.734 & 1.696 & 2.174 & 2.353 & 2.457  \\
        \hline \hline
    \end{tabular}}
    \label{f-c}
\end{table*}
Andersson and Kokkotas \cite{Kokkotas1999} first established the correlation between $C$ and $f$-mode frequency. Here, we calculate the $C$-$f$ relations for NSs in EMSG, using the approximate formula obtained through least-squares fitting
\begin{equation}
\label{eq:C-f_fitting}
    C = \sum_{n=0}^{n=4} b_n (\Bar{\omega})^n \, .
\end{equation}
The compactness is represented as a function of the normalized $f$-mode frequency ($\Bar{\omega}$) in Fig. \ref{cf1} with $\alpha (10^{-38}) = -5.01,  0 \& +5.01$.
for NSs. The coefficients ($b_n$) with $\chi_r^2-$ error are enumerated in Table \ref{f-c}. The magnitude of some $b_n$ values increases and decreases with the change in $\alpha$, suggesting that the fitting is more robust for the isotropic ($\alpha=0$ (GR)) case. Additionally, $\chi_r^2$ increases by including the positive $\alpha$ parameters and decreases by including the negative $\alpha$ parameters. Thus, including $\alpha$ (positive or negative) diminishes the EOS insensitivity. One of the primary applications of $C$-$f$ URs involves determining $M$ and $R$ based on the analysis of observed mode data, as articulated by Andersson and Kokkotas \cite{Andersson_1998}. 

\subsection{Comparison Study}
\label{comp}

We constrain the canonical $f$-mode frequency for GW170817 \cite{GW170817} and GW190814 \cite{GW190814} events across five different values of $\alpha$ parameters, as outlined in Table \ref{tab:canonical_f-mode}. The canonical $f$-mode frequency is also compared with previous studies, focusing on isotropic ($\alpha=0$ (GR)) NS, as listed in Table \ref{tab:comparison}. Notably, the $f$-mode frequency obtained in this study is approximately 30-35\% more than the findings of Chan {\it et al.} \cite{Chan_2014}, Pradhan {\it et al.} \cite{Bikram_2023}, Sotani and Kumar \cite{Sotani_2021}. This difference in the $f$-mode was anticipated, given that the aforementioned authors (except Mohanty {\it et al.} \cite{Mohanty_2024}, who used Cowling approximation) employed a full-GR formalism for their $f$-mode calculations, in contrast to our use of the Cowling approximation in this study.
\begin{table*}[h!]
\centering
\caption{The canonical $f$-mode frequency ($f_{1.4}$ in kHz) inferred from GW170817 and GW190814 data using $f$-Love UR obtained in different literature for isotropic ( here $\alpha=0$ (GR)) NS.}
\renewcommand{\arraystretch}{1.3}
\scalebox{0.8}{
\begin{tabular}{cccc}
    \hline \hline & \multicolumn{1}{c}{$\mathrm{GW170817}$} & & \multicolumn{1}{c}{$\mathrm{GW190814}$} \\
    \hline
     Ref. & $f_{1.4}$ & & $f_{1.4}$ \\ 
    \hline \vspace{-3mm} \\
    \parbox[c]{0.35\linewidth}{\centering Chan {\it et al.}  \\ \cite{Chan_2014}} & $2.120 ^{+0.445} _ {-0.446}$ & & $1.652 ^{+0.111} _ {-0.130}$ \vspace{1.5mm} \\

    \parbox[c]{0.35\linewidth}{\centering Pradhan  {\it et al.} \\  \cite{Bikram_2023}} & $2.120 ^{+0.444} _ {-0.445}$ & & $1.653 ^{+0.111} _ {-0.130}$ \vspace{1.5mm} \\
    
    \parbox[c]{0.35\linewidth}{\centering Sotani and Kumar  \\ \cite{Sotani_2021}} & $2.124 ^{+0.440} _ {-0.446}$ & & $1.656 ^{+0.112} _ {-0.132}$ \vspace{1.5mm} \\

    \parbox[c]{0.35\linewidth}{\centering
     Mohanty {\it et al.} \\ 
     \cite{Mohanty_2024}} & $2.606^{+0.457} _ {-0.484}$ & & $2.097^{+0.124} _ {-0.149}$ \vspace{2mm} \\ 
     This Work  & $2.659^{+0.458} _ {-0.491}$ & & $2.143^{+0.127} _ {-0.153}$ \vspace{2mm} \\ 
    \hline \hline
\end{tabular}}
\label{tab:comparison}
\end{table*}
\section{Results and Discussion} 
\label{RD}
In this work, we considered stiff, intermediate, and soft EOSs. We varied the $\alpha$ parameter from negative to positive values to investigate how this impacts key physical quantities such as sound speed, mass, radius, and $f$-mode frequency.
\begin{figure*}[h!]
    \includegraphics[width=0.45\linewidth]{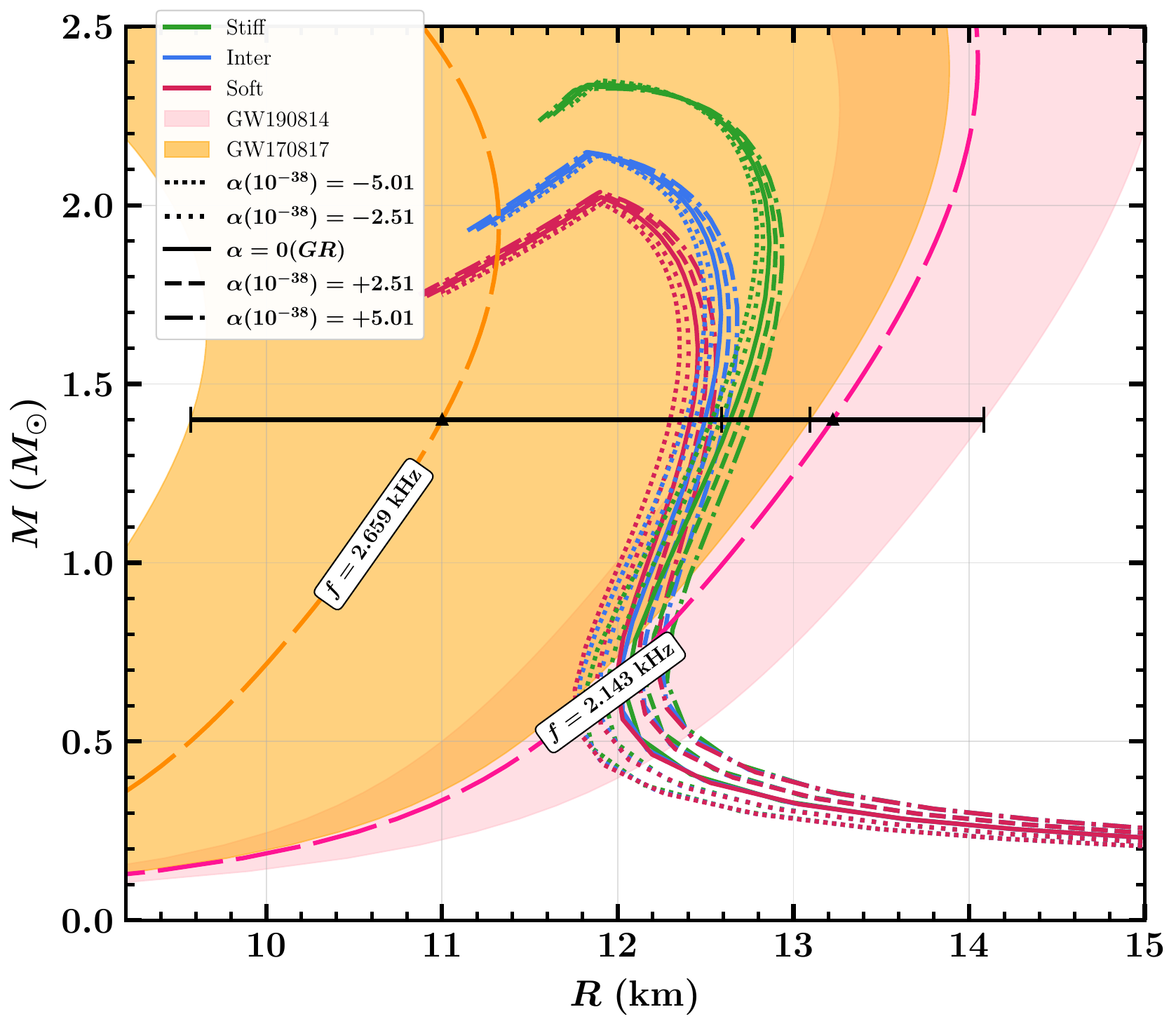}
    \includegraphics[width=0.55\textwidth]{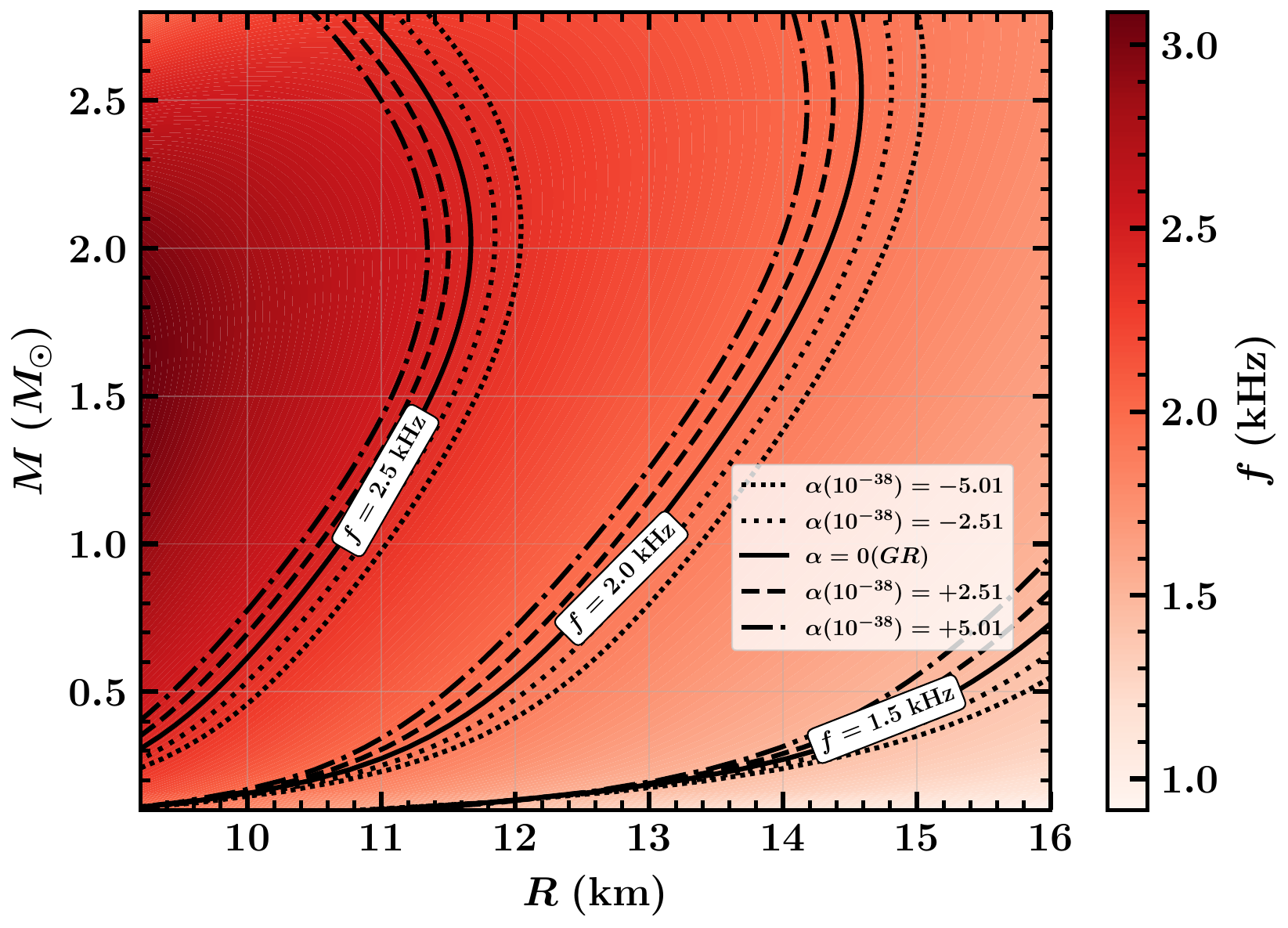}
    \caption{\textit{Left:} The orange and pink coloured MR bands correspond to limits on NS’s mass and radius imposed by isotropic ($\alpha=0$ (GR)) $C$-$f$ UR for the canonical $f$-mode frequency ($f_{1.4}$) that was obtained through $f$-Love UR
with the help of tidal deformability constraints ($\Lambda_{1.4}$) of GW170817 \cite{GW170817,Bharat_and_Landry} and GW190814 \cite{GW190814} events,
respectively. The horizontal error bars illustrate the radius limits for a canonical mass NS imposed by
the frequency bands for respective events.
\textit{Right:}  The frequency distribution contour plot across the M–R parameter space. This distribution is imposed by the $C$-$f$ UR for $\alpha=0$.
    }
    \label{mr}
\end{figure*}
\begin{table*}[h!]
\caption{Maximum mass (M$_{max}$), corresponding radius (R$_{max}$), radius at 1.4M$_\odot$ ($R_{1.4}$) and $f$-mode frequency at 1.4M$_\odot$ ($f_{1.4}$) of the NS for the EOSs corresponding to \textbf{Stiff}, \textbf{Inter}, \textbf{Soft}, calculated with the variation of $\alpha$ parameters.}
\centering
\setlength{\tabcolsep}{5pt}
\renewcommand{\arraystretch}{1}
\scalebox{0.7}{\begin{tabular}{ccccccc}
\hline
\hline\noalign{\smallskip} 
{\begin{tabular}[c]{@{}l@{}}EOS models\end{tabular}}&
{\begin{tabular}[c]{@{}l@{}}Alpha parameter \\  ($\alpha (10^{-38})$) (cm$^3$/erg)\end{tabular}}&
{\begin{tabular}[c]{@{}l@{}}Maximum Mass\\  ($M_{max}$) (M$_\odot$)\end{tabular}}&
{\begin{tabular}[c]{@{}l@{}}Maximum Radius\\($R_{max}$) (Km)\end{tabular}}&
{\begin{tabular}[c]{@{}l@{}}Radius at 1.4M$_\odot$\\($R_{1.4}$) (Km)\end{tabular}}&
{\begin{tabular}[c]{@{}l@{}}$f$-mode freq. at 1.4M$_\odot$ \\($f_{1.4})$(kHz)\end{tabular}}
\\
\hline
\hline\noalign{\smallskip} 
\textbf{Stiff} \\
\hline\noalign{\smallskip} 
& $-5.01$   & 2.35  & 11.90  & 12.53  & 2.29   \\
& $-2.51$   & 2.34  & 11.87  & 12.59  & 2.23   \\
& 0         & 2.34  & 11.89  & 12.64  & 2.18   \\ 
& $+2.51$   & 2.33  & 11.96  & 12.69  & 2.14   \\
& $+5.01$   & 2.33  & 11.96  & 12.75  & 2.09   \\ 
\hline\noalign{\smallskip} 
\textbf{Inter} \\
\hline\noalign{\smallskip}
& $-5.01$   & 2.14  & 11.88  & 12.39  & 2.32   \\
& $-2.51$   & 2.14  & 11.87  & 12.44  & 2.27   \\
& 0         & 2.14  & 11.85  & 12.50  & 2.22   \\ 
& $+2.51$   & 2.15  & 11.84  & 12.55  & 2.17   \\
& $+5.01$   & 2.15  & 11.82  & 12.61  & 2.12   \\ 
 \hline\noalign{\smallskip} 
\textbf{Soft} \\
\hline\noalign{\smallskip}
& $-5.01$   & 2.01  & 11.91  & 12.30  & 2.34   \\
& $-2.51$   & 2.01  & 11.91  & 12.36  & 2.29   \\
& 0         & 2.02  & 11.90  & 12.41  & 2.23   \\ 
& $+2.51$   & 2.03  & 11.90  & 12.47  & 2.19   \\
& $+5.01$   & 2.04  & 11.90  & 12.53  & 2.14   \\ 
\noalign{\smallskip}\hline
\hline
\end{tabular}}
\label{table:f_var}
\end{table*}
\\
Fig. \ref{EOS} (Left panel) illustrates the variation of pressure ($P$) as a function of energy density ($\mathcal{E}$) for three distinct EOS—Stiff (green), Intermediate (blue), and Soft (red). The horizontal plateaus observed in the pressure curves signify first-order phase transitions, during which the pressure remains constant as the energy density increases. For the Stiff EOS, the phase transition occurs between energy densities of approximately 1100 MeV/fm$^3$ and 2110 MeV/fm$^3$, with a constant pressure of about 445 MeV/fm$^3$. The Intermediate EOS experiences a phase transition across the range of 930 MeV/fm$^3$ to 1750 MeV/fm$^3$, maintaining a constant pressure around 300 MeV/fm$^3$. In contrast, the Soft EOS undergoes a phase transition from approximately 850 MeV/fm$^3$ to 1500 MeV/fm$^3$, with a constant pressure of around 230 MeV/fm$^3$. Vertical dotted lines demarcate the phase transition boundaries for each EOS, providing a clear visual indication of the initiation and termination of the transition. The Stiff EOS experiences the phase transition at the highest energy densities and pressures, followed by the Intermediate and Soft EOSs, highlighting the distinctive characteristics of these models. Subsequent to the transition, the pressure increases sharply once more, indicating a return to the typical behaviour of the EOS. This depiction facilitates a clear comparison of the phase transition regions and behaviours across the different EOS models.
\\
In the Right panel of Fig. \ref{EOS}, the variation of the sound speed squared ($c_s^2$) from the centre to the surface of NSs is depicted for three distinct EOSs: Stiff (green), Intermediate (blue), and Soft (red), under five different $\alpha$ parameter values: $\alpha = -5.01 \times 10^{-38}$, $\alpha = -2.51 \times 10^{-38}$, $\alpha = 0$, $\alpha = +2.51 \times 10^{-38}$, and $\alpha = +5.01 \times 10^{-38}$. The plot demonstrates that for all $\alpha$ values, the sound speed squared, $c_s^2$, satisfied the causality limit, ensuring that $0 \leq c_s^2 \leq 1$. The conformal limit $c_s^2 = 1/3$ is indicated by a golden dashed line, serving as a reference for EOS stiffness. In the core regions of the NS, $c_s^2$ attains values ranging from 0.40 to 0.60, depending on both the EOS and $\alpha$, with the Stiff EOS exhibiting the highest values close to the centre. As the radius increases, the sound speed shows a consistent decline, reaching values below 0.2 near the star's surface at approximately 10-12 km for all EOS models. For larger $\alpha$ values, particularly $\alpha = +5.01 \times 10^{-38}$, the curves persist at slightly elevated levels in comparison to the Right $\alpha$ values, especially in the central regions (refer to the zoomed plot). The figure also shows phase transitions, indicated by arrows, wherein the squared sound speed exhibits a noticeable shift. These phase transitions are localized closer to the centre, specifically between 1.5 km and 3 km from the core, with the Stiff EOS (green) presenting a transition approximately at $r \approx 1.8$ km, the Intermediate EOS (blue) at $r \approx 2.2$ km, and the Soft EOS (red) at $r \approx 2.6$ km. These transitions signify alterations in the star's internal composition, potentially from hadronic matter to a quark or mixed phase. The transitions are characterized as smooth, congruent with second-order phase transitions, and occur within conformal and causality boundaries, thereby ensuring that no physical constraints are violated. The gradual decline in $c_s^2$ subsequent to the transition further validates the consistency of the EOS models. The radial profiles for various $\alpha$ values exhibit similar behaviours across the EOS types, with minor variations at small radii and slight differences in the sound speed near the surface for the stiffest EOSs. This behaviour facilitates the comparison of the effects of $\alpha$ and the selected EOS on the star's sound speed profile, thereby reaffirming the physical plausibility of the models across different parameter values.
\\
The left panel of Fig. \ref{f_mode} represents the variation of non-radial $f$-mode frequency with the mass of NSs. When we are calculating the frequency in the EMSG, we vary the $\alpha$ parameter with five different negative to positive values. We have seen that, the $f$-mode frequency decreases with positive $\alpha$ and increases with negative $\alpha$. The verticle black-dashed line is the 1.4 M$_\odot$ line so we can see the variation of $f_{1.4}$ (frequency at 1.4 M$_\odot$) for different $\alpha$ values for different EOSs. In the last column of Table \ref{table:f_var} we have noted the variation of $f$-mode frequency at 1.4M$_\odot$ ($f_{1.4}$) of the NS. The right panel of Fig. \ref{f_mode} represents the variation of non-radial $f$-mode frequency with the central energy density of NSs
\\
Then, we established the $f$-Love relation, the relation between the $f$-mode frequency and tidal Love number. In Fig. \ref{flr1}, we have presented this relation and the best-fit results, from which we have calculated the canonical frequencies corresponding to the tidal Love number from the two GW events. Using this $f$-Love relation, we have also established the $C-f$ relation, the relation between the compactness ($C=M/R$) and the $f$-mode oscillation frequency, presented in Fig. \ref{cf1}. After that, we used these two relations and constrained the $f$-mode oscillation frequency in the $M-R$ plane. We have depicted those results in Fig. \ref{mr}. Also in this Fig. \ref{mr} (Left panel), we have shown how the mass and radius of the NSs vary with the $\alpha$ parameter and those values are noted in Table \ref{table:f_var}. For Soft and Intermediate EOSs, if we consider the positive values of $\alpha$, the maximum mass of the NS increase and the radius corresponding to maximum mass decreases, that's why the compactness ($C=M/R$) of the NS increases, i.e. we cant easily deformed the NS, and for that reason the oscillation frequency decrease. For negative values of $\alpha$, the opposite thing happens. But for Stiff EOS, the phase transition happens before attaining the maximum mass, so the maximum mass and corresponding radius vary abruptly.
\\
For a given $f$-mode frequency, the $C$-$f$ UR has been used to derive an $M$-$R$ relation. This constrained relation, incorporating uncertainties as standard deviations in the UR, generates $M$-$R$ bands, shown in the Left panel of Fig. \ref{mr}. The orange $M$-$R$ band represents the region where NSs are expected to have a frequency of $f = 2.659^{+0.458}_{-0.491}$ kHz, with the solid dashed line indicating $f = 2.659$ kHz. Similarly, the pink band corresponds to a frequency of $f = 2.143^{+0.127}_{-0.153}$ kHz, with the solid dashed line marking $f = 2.143$ kHz. These frequency constraints align with the canonical $f$-mode frequency bounds determined in this study for isotropic ($\alpha = 0$ (GR)) NSs in the GW170817 \cite{GW170817} and GW190814 \cite{GW190814} events. The Right panel of Fig. \ref{mr} displays the distribution of $f$-mode frequencies over the $M$-$R$ parameter space for isotropic ($\alpha = 0$ (GR)) NSs based on the $C$-$f$ UR. The black solid line highlights specific mass-radius combinations for isotropic stars predicted to exhibit these frequencies according to the $C$-$f$ UR. Additionally, variations in these $M$-$R$ lines due to the inclusion of the $\alpha$ parameter are illustrated in the figure.
NSs with frequencies $f<1.5$ kHz are situated in the low compactness region and undergo minimal changes in mass and radius due to the inclusion of $\alpha$. By observing $M$-$R$ lines as shown in the Right panel of Fig. \ref{mr}, we conclude that for a constant mass NS having a fixed frequency of $f \geq 1.5$ kHz, the radius tends to decrease with the presence of positive $\alpha$ and increase for negative $\alpha$, affecting the compactness of the star to maintain its natural frequency. 


\section{Summary \& Conclusions}
\label{conclusion}
In this study, we have explored the macroscopic properties like mass, radius and $f$-mode frequency for non-radial oscillation of NSs in EMSG. Our main objective of this study is to constrain the $f$-mode oscillation frequency using universal relations. For this study, we have considered three different kinds of EOSs: Stiff, Intermediate and Soft.
\\
So, we considered the EMSG and modified every equation required to calculate the properties of NSs. Initially, we considered a free parameter $\alpha$ (related to Energy-momentum square, $\alpha T_{\mu\nu}T^{\mu\nu}$), which we can vary based upon its limit constrained for studying the NSs in EMSG. Then, solve the Hydrostatic equilibrium in EMSG and get the modified TOV equations in terms of $\alpha$. So, we solve the TOV equations using the boundary conditions and with those aforementioned EOSs while varying the $\alpha$ parameter with negative and positive values. Here in this work, we have considered five different values of $\alpha$. After solving TOV equations, we get the mass and radius data for NSs corresponding to five different $\alpha$ values.
\\
Then, we focus on the non-radial oscillation and modify them with the effective energy and pressure density we got from previous equations. For this study, we concentrate only on the $f$- mode oscillation frequency.
\\
After that, we forward our study towards URs. We considered two relations, one between $f$-mode frequency and the tidal love number ($f$-Love), and another one between compactness ($C$) and $f$-mode frequency ($C-f$). Using these two relations, we successfully put the constraint on the $f$-mode frequency in the $M-R$ plane. For this, we have taken the source data from two well-known GW events, GW170817 and GW190814. In the end, we did a comparison study with some previous work; those who had calculated the $f$-mode frequency from the two known GW events, and found that our results are well satisfied with those works.
\\
EMSG introduces significant modifications to the macroscopic properties of NSs compared to GR. The $\alpha$ parameter in EMSG alters the stiffness of the EOS, impacting the mass-radius ($M$-$R$) relation, tidal deformability, and oscillation frequencies. Positive $\alpha$ values lead to stiffer EOSs, while negative values result in softer EOSs. These effects extend to universal relations, such as those connecting $f$-mode frequencies to compactness, which differ notably from GR predictions. This framework allows for deviations in NS structure and oscillation frequencies, potentially testing alternative theories of gravity through GW observations like GW170817 and GW190814. So we can say that EMSG offers a compelling approach to studying strong-field gravity and dense matter, with future GW observations assured of its existence.


\section{Acknowledgments}








I would like to thank my supervisor Dr. Bharat Kumar for his fruitful discussions and the needful suggestions. I also want to thank Mr. Sailesh Ranjan Mohanty and Mr. Nishant Singh for their help with the computations. Author acknowledges partial support from the Department of Science and Technology, Government of India, with grant no. CRG/2021/000101.

\section*{Declarations}


\begin{itemize}
\item \textbf{Funding} Department of Science and Technology, Government of India.
\item \textbf{Conflict of interest/Competing interests} The author declares no Conflict of interest.
\item \textbf{Data availability} Data sets generated during the current study are available from the corresponding author
on reasonable request.
\end{itemize}

\bibliography{sn-bibliography}

\begin{thebibliography}{10}
\providecommand{\url}[1]{{#1}}
\providecommand{\urlprefix}{URL }
\providecommand{\doi}[1]{\url{https://doi.org/#1}}
\bibcommenthead

\bibitem{doi:10.1126/science.aap9811}
D.A. Coulter, et~al., Swope supernova survey 2017a (sss17a), the optical counterpart to a gravitational wave source.
\newblock Science \textbf{358}(6370), 1556--1558 (2017).
\newblock \doi{10.1126/science.aap9811}.
\newblock \urlprefix\url{https://www.science.org/doi/abs/10.1126/science.aap9811}.
\newblock {\href{https://arxiv.org/abs/https://www.science.org/doi/pdf/10.1126/science.aap9811}{{https://www.science.org/doi/pdf/10.1126/science.aap9811}}}

\bibitem{Abbott_2020}
B.P. Abbott, Gw190425: Observation of a compact binary coalescence with total mass 3.4 solar mass.
\newblock The Astrophysical Journal Letters \textbf{892}(1), L3 (2020).
\newblock \doi{10.3847/2041-8213/ab75f5}.
\newblock \urlprefix\url{https://dx.doi.org/10.3847/2041-8213/ab75f5}

\bibitem{PhysRevX.9.011001}
B.P. Abbott, Properties of the binary neutron star merger gw170817.
\newblock Phys. Rev. X \textbf{9}, 011001 (2019).
\newblock \doi{10.1103/PhysRevX.9.011001}.
\newblock \urlprefix\url{https://link.aps.org/doi/10.1103/PhysRevX.9.011001}

\bibitem{Fragione_2021}
G.~Fragione, Black-hole–neutron-star mergers are unlikely multimessenger sources.
\newblock The Astrophysical Journal Letters \textbf{923}(1), L2 (2021).
\newblock \doi{10.3847/2041-8213/ac3bcd}.
\newblock \urlprefix\url{http://dx.doi.org/10.3847/2041-8213/ac3bcd}

\bibitem{sgrb1}
P.~D'Avanzo, Short gamma-ray bursts: A review.
\newblock Journal of High Energy Astrophysics \textbf{7}, 73--80 (2015).
\newblock \doi{https://doi.org/10.1016/j.jheap.2015.07.002}.
\newblock \urlprefix\url{https://www.sciencedirect.com/science/article/pii/S2214404815000415}.
\newblock Swift 10 Years of Discovery, a novel approach to Time Domain Astronomy

\bibitem{sgrb2}
N.~Jordana-Mitjans, C.G. Mundell, C.~Guidorzi, R.J. Smith, E.~Ramírez-Ruiz, B.D. Metzger, S.~Kobayashi, A.~Gomboc, I.A. Steele, M.~Shrestha, M.~Marongiu, A.~Rossi, B.~Rothberg, A short gamma-ray burst from a protomagnetar remnant.
\newblock The Astrophysical Journal \textbf{939}(2), 106 (2022).
\newblock \doi{10.3847/1538-4357/ac972b}.
\newblock \urlprefix\url{https://dx.doi.org/10.3847/1538-4357/ac972b}

\bibitem{Abbott_2017}
B.P. Abbott, R.~Abbott, T.D. Abbott, Multi-messenger observations of a binary neutron star merger*.
\newblock The Astrophysical Journal Letters \textbf{848}(2), L12 (2017).
\newblock \doi{10.3847/2041-8213/aa91c9}.
\newblock \urlprefix\url{https://dx.doi.org/10.3847/2041-8213/aa91c9}

\bibitem{GW170817}
B.P. Abbott, R.~Abbott, A.~et~al., Gw170817: Observation of gravitational waves from a binary neutron star inspiral.
\newblock Phys. Rev. Lett. \textbf{119}, 161101 (2017).
\newblock \doi{10.1103/PhysRevLett.119.161101}.
\newblock \urlprefix\url{https://link.aps.org/doi/10.1103/PhysRevLett.119.161101}

\bibitem{CAPOZZIELLO2011167}
S.~Capozziello, M.~{De Laurentis}, Extended theories of gravity.
\newblock Physics Reports \textbf{509}(4), 167--321 (2011).
\newblock \doi{https://doi.org/10.1016/j.physrep.2011.09.003}.
\newblock \urlprefix\url{https://www.sciencedirect.com/science/article/pii/S0370157311002432}

\bibitem{OLMO20201}
G.J. Olmo, D.~Rubiera-Garcia, A.~Wojnar, Stellar structure models in modified theories of gravity: Lessons and challenges.
\newblock Physics Reports \textbf{876}, 1--75 (2020).
\newblock \doi{https://doi.org/10.1016/j.physrep.2020.07.001}.
\newblock \urlprefix\url{https://www.sciencedirect.com/science/article/pii/S0370157320302507}.
\newblock Stellar structure models in modified theories of gravity: Lessons and challenges

\bibitem{PhysRevD.98.024031}
N.~Nari, M.~Roshan, Compact stars in energy-momentum squared gravity.
\newblock Phys. Rev. D \textbf{98}, 024031 (2018).
\newblock \doi{10.1103/PhysRevD.98.024031}.
\newblock \urlprefix\url{https://link.aps.org/doi/10.1103/PhysRevD.98.024031}

\bibitem{PhysRevD.84.024020}
T.~Harko, F.S.N. Lobo, S.~Nojiri, S.D. Odintsov, $f(r,t)$ gravity.
\newblock Phys. Rev. D \textbf{84}, 024020 (2011).
\newblock \doi{10.1103/PhysRevD.84.024020}.
\newblock \urlprefix\url{https://link.aps.org/doi/10.1103/PhysRevD.84.024020}

\bibitem{EMSG_OAkarsu}
Özgür Akarsu~et.al., Constraint on energy-momentum squared gravity from neutron stars and its cosmological implications.
\newblock Phys. Rev. D \textbf{97}, 124017 (2018).
\newblock \doi{10.1103/PhysRevD.97.124017}.
\newblock \urlprefix\url{https://link.aps.org/doi/10.1103/PhysRevD.97.124017}

\bibitem{Nari}
N.~Nari, M.~Roshan, Compact stars in energy-momentum squared gravity.
\newblock Phys. Rev. D \textbf{98}, 024031 (2018).
\newblock \doi{10.1103/PhysRevD.98.024031}.
\newblock \urlprefix\url{https://link.aps.org/doi/10.1103/PhysRevD.98.024031}

\bibitem{Akarsu}
O.~Akarsu, J.D. Barrow, N.M. Uzun, Screening anisotropy via energy-momentum squared gravity: $\mathrm{\ensuremath{\Lambda}}\mathrm{CDM}$ model with hidden anisotropy.
\newblock Phys. Rev. D \textbf{102}, 124059 (2020).
\newblock \doi{10.1103/PhysRevD.102.124059}.
\newblock \urlprefix\url{https://link.aps.org/doi/10.1103/PhysRevD.102.124059}

\bibitem{Katırcı2014}
N.~Katirci, M.~Kavuk, $f(r, t_{\mu\nu} t^{\mu\nu})$ gravity and cardassian-like expansion as one of its consequences.
\newblock The European Physical Journal Plus \textbf{129}(8), 163 (2014).
\newblock \doi{10.1140/epjp/i2014-14163-6}.
\newblock \urlprefix\url{https://doi.org/10.1140/epjp/i2014-14163-6}

\bibitem{Nazari}
E.~Nazari, Light bending and gravitational lensing in energy-momentum-squared gravity.
\newblock Phys. Rev. D \textbf{105}, 104026 (2022).
\newblock \doi{10.1103/PhysRevD.105.104026}.
\newblock \urlprefix\url{https://link.aps.org/doi/10.1103/PhysRevD.105.104026}

\bibitem{Nazari1}
E.~Nazari, M.~Roshan, I.~De~Martino, Constraining energy-momentum-squared gravity by binary pulsar observations.
\newblock Phys. Rev. D \textbf{105}, 044014 (2022).
\newblock \doi{10.1103/PhysRevD.105.044014}.
\newblock \urlprefix\url{https://link.aps.org/doi/10.1103/PhysRevD.105.044014}

\bibitem{Fazlollahi2023}
H.R. Fazlollahi, Energy--momentum squared gravity and late-time universe.
\newblock The European Physical Journal Plus \textbf{138}(3), 211 (2023).
\newblock \doi{10.1140/epjp/s13360-023-03723-w}.
\newblock \urlprefix\url{https://doi.org/10.1140/epjp/s13360-023-03723-w}

\bibitem{Danarianto2023}
M.D. Danarianto, A.~Sulaksono, Towards precise constraints in modified gravity: bounds on alternative coupling gravity using white dwarf mass-radius measurements.
\newblock The European Physical Journal C \textbf{83}(6), 463 (2023).
\newblock \doi{10.1140/epjc/s10052-023-11644-2}.
\newblock \urlprefix\url{https://doi.org/10.1140/epjc/s10052-023-11644-2}

\bibitem{Chandrasekhar_1964}
S.~{Chandrasekhar}, {The Dynamical Instability of Gaseous Masses Approaching the Schwarzschild Limit in General Relativity.}
\newblock apj \textbf{140}, 417 (1964).
\newblock \doi{10.1086/147938}

\bibitem{Chanmugam_1977}
G.~{Chanmugam}, {Radial oscillations of zero-temperature white dwarfs and neutron stars below nuclear densities.}
\newblock apj \textbf{217}, 799--808 (1977).
\newblock \doi{10.1086/155627}

\bibitem{Kokkotas_2001}
K.D. Kokkotas, J.~Ruoff, Radial oscillations of relativistic stars.
\newblock Astronomy \& Astrophysics \textbf{366}(2), 565--572 (2001).
\newblock \doi{10.1051/0004-6361:20000216}.
\newblock \urlprefix\url{https://doi.org/10.1051%2F0004-6361%3A20000216}

\bibitem{mcdermott_1988}
P.N. {McDermott}, H.M. {van Horn}, C.J. {Hansen}, {Nonradial Oscillations of Neutron Stars}.
\newblock apj \textbf{325}, 725 (1988).
\newblock \doi{10.1086/166044}

\bibitem{Kunjipurayil_2022}
A.~Kunjipurayil, T.~Zhao, B.~Kumar, B.K. Agrawal, M.~Prakash, Impact of the equation of state on $f$- and $p$- mode oscillations of neutron stars.
\newblock Phys. Rev. D \textbf{106}, 063005 (2022).
\newblock \doi{10.1103/PhysRevD.106.063005}.
\newblock \urlprefix\url{https://link.aps.org/doi/10.1103/PhysRevD.106.063005}

\bibitem{Zhao_2022}
T.~Zhao, J.M. Lattimer, Universal relations for neutron star $f$-mode and $g$-mode oscillations.
\newblock Phys. Rev. D \textbf{106}, 123002 (2022).
\newblock \doi{10.1103/PhysRevD.106.123002}.
\newblock \urlprefix\url{https://lineutron stars with short-range correlation and admixed dark matternk.aps.org/doi/10.1103/PhysRevD.106.123002}

\bibitem{Sotani_2021}
H.~Sotani, B.~Kumar, Universal relations between the quasinormal modes of neutron star and tidal deformability.
\newblock Phys. Rev. D \textbf{104}, 123002 (2021).
\newblock \doi{10.1103/PhysRevD.104.123002}.
\newblock \urlprefix\url{https://link.aps.org/doi/10.1103/PhysRevD.104.123002}

\bibitem{Finn_1987}
L.S. Finn, {g-modes in zero-temperature neutron stars}.
\newblock Mon. Not. R. Astron. Soc. \textbf{227}(2), 265--293 (1987).
\newblock \doi{10.1093/mnras/227.2.265}

\bibitem{Haskell_2014}
B.~Haskell, K.~Glampedakis, N.~Andersson, {A new mechanism for saturating unstable r modes in neutron stars}.
\newblock Mon. Not. R. Astron. Soc. \textbf{441}(2), 1662--1668 (2014).
\newblock \doi{10.1093/mnras/stu535}

\bibitem{Benhar_1999}
O.~Benhar, E.~Berti, V.~Ferrari, {The imprint of the equation of state on the axial w-modes of oscillating neutron stars}.
\newblock Mon. Not. R. Astron. Soc. \textbf{310}(3), 797--803 (1999).
\newblock \doi{10.1046/j.1365-8711.1999.02983.x}

\bibitem{Tianqi_2022}
T.~Zhao, C.~Constantinou, P.~Jaikumar, et~al., Quasinormal $g$ modes of neutron stars with quarks.
\newblock Phys. Rev. D \textbf{105}, 103025 (2022).
\newblock \doi{10.1103/PhysRevD.105.103025}.
\newblock \urlprefix\url{https://link.aps.org/doi/10.1103/PhysRevD.105.103025}

\bibitem{Shibagaki_2020}
S.~Shibagaki, T.~Kuroda, K.~Kotake, et~al., {A new gravitational-wave signature of low-T/|W| instability in rapidly rotating stellar core collapse}.
\newblock Monthly Notices of the Royal Astronomical Society: Letters \textbf{493}(1), L138--L142 (2020).
\newblock \doi{10.1093/mnrasl/slaa021}.
\newblock \urlprefix\url{https://doi.org/10.1093/mnrasl/slaa021}

\bibitem{Sotani_2011}
H.~Sotani, N.~Yasutake, T.~Maruyama, et~al., Signatures of hadron-quark mixed phase in gravitational waves.
\newblock Phys. Rev. D \textbf{83}, 024014 (2011).
\newblock \doi{10.1103/PhysRevD.83.024014}.
\newblock \urlprefix\url{https://link.aps.org/doi/10.1103/PhysRevD.83.024014}

\bibitem{Flores_2014}
C.V. Flores, G.~Lugones, Discriminating hadronic and quark stars through gravitational waves of fluid pulsation modes.
\newblock Classical and Quantum Gravity \textbf{31}(15), 155002 (2014).
\newblock \doi{10.1088/0264-9381/31/15/155002}.
\newblock \urlprefix\url{https://doi.org/10.1088/0264-9381/31/15/155002}

\bibitem{Andersson_1998}
N.~Andersson, K.D. Kokkotas, {Towards gravitational wave asteroseismology}.
\newblock Monthly Notices of the Royal Astronomical Society \textbf{299}(4), 1059--1068 (1998).
\newblock \doi{10.1046/j.1365-8711.1998.01840.x}.
\newblock \urlprefix\url{https://doi.org/10.1046/j.1365-8711.1998.01840.x}

\bibitem{Kent_yagi_2013}
K.~Yagi, N.~Yunes, I-love-q relations in neutron stars and their applications to astrophysics, gravitational waves, and fundamental physics.
\newblock Phys. Rev. D \textbf{88}, 023009 (2013).
\newblock \doi{10.1103/PhysRevD.88.023009}.
\newblock \urlprefix\url{https://link.aps.org/doi/10.1103/PhysRevD.88.023009}

\bibitem{Mohanty_2024}
S.R. Mohanty, S.~Ghosh, P.~Routaray, H.~Das, B.~Kumar, The impact of anisotropy on neutron star properties: insights from i-f-c universal relations.
\newblock Journal of Cosmology and Astroparticle Physics \textbf{2024}(03), 054 (2024).
\newblock \doi{10.1088/1475-7516/2024/03/054}.
\newblock \urlprefix\url{https://dx.doi.org/10.1088/1475-7516/2024/03/054}

\bibitem{Rahmansyah}
A.~Rahmansyah, D.~Purnamasari, R.~Kurniadi, A.~Sulaksono, Generalized tolman-oppenheimer-volkoff model and neutron stars.
\newblock Phys. Rev. D \textbf{106}, 084042 (2022).
\newblock \doi{10.1103/PhysRevD.106.084042}.
\newblock \urlprefix\url{https://link.aps.org/doi/10.1103/PhysRevD.106.084042}

\bibitem{HQPT}
T.~Demircik, C.~Ecker, M.~J\"arvinen, Dense and hot qcd at strong coupling.
\newblock Phys. Rev. X \textbf{12}, 041012 (2022).
\newblock \doi{10.1103/PhysRevX.12.041012}.
\newblock \urlprefix\url{https://link.aps.org/doi/10.1103/PhysRevX.12.041012}

\bibitem{EMSG_NAlam}
N.~Alam, S.~Pal, A.~Rahmansyah, A.~Sulaksono, Impact of modified gravity theory on neutron star and nuclear matter properties.
\newblock Phys. Rev. D \textbf{109}, 083007 (2024).
\newblock \doi{10.1103/PhysRevD.109.083007}.
\newblock \urlprefix\url{https://link.aps.org/doi/10.1103/PhysRevD.109.083007}

\bibitem{Faraoni}
V.~Faraoni, Lagrangian description of perfect fluids and modified gravity with an extra force.
\newblock Phys. Rev. D \textbf{80}, 124040 (2009).
\newblock \doi{10.1103/PhysRevD.80.124040}.
\newblock \urlprefix\url{https://link.aps.org/doi/10.1103/PhysRevD.80.124040}

\bibitem{PhysRevD.109.104055}
O.~Akarsu, M.~Bouhmadi-L\'opez, N.~Kat\ifmmode \imath \else \i \fi{}rc\ifmmode \imath \else~\i \fi{}, E.~Nazari, M.~Roshan, N.M. Uzun, Equivalence of matter-type modified gravity theories to general relativity with nonminimal matter interaction.
\newblock Phys. Rev. D \textbf{109}, 104055 (2024).
\newblock \doi{10.1103/PhysRevD.109.104055}.
\newblock \urlprefix\url{https://link.aps.org/doi/10.1103/PhysRevD.109.104055}

\bibitem{Wald:1984rg}
R.M. Wald, \emph{{General Relativity}} (Chicago Univ. Pr., Chicago, USA, 1984).
\newblock \doi{10.7208/chicago/9780226870373.001.0001}

\bibitem{Schwarzschild:1916uq}
K.~Schwarzschild, {On the gravitational field of a mass point according to Einstein's theory}.
\newblock Sitzungsber. Preuss. Akad. Wiss. Berlin (Math. Phys. ) \textbf{1916}, 189--196 (1916).
\newblock {\href{https://arxiv.org/abs/physics/9905030}{{arXiv:physics/9905030}}}

\bibitem{PhysRevD.66.104002}
N.~Andersson, G.L. Comer, D.~Langlois, Oscillations of general relativistic superfluid neutron stars.
\newblock Phys. Rev. D \textbf{66}, 104002 (2002).
\newblock \doi{10.1103/PhysRevD.66.104002}.
\newblock \urlprefix\url{https://link.aps.org/doi/10.1103/PhysRevD.66.104002}

\bibitem{Cowling_1941}
T.G. Cowling, {The Non-radial Oscillations of Polytropic Stars}.
\newblock Monthly Notices of the Royal Astronomical Society \textbf{101}(8), 367--375 (1941).
\newblock \doi{10.1093/mnras/101.8.367}.
\newblock \urlprefix\url{https://doi.org/10.1093/mnras/101.8.367}

\bibitem{PhysRevD.83.024014}
H.~Sotani, N.~Yasutake, T.~Maruyama, T.~Tatsumi, Signatures of hadron-quark mixed phase in gravitational waves.
\newblock Phys. Rev. D \textbf{83}, 024014 (2011).
\newblock \doi{10.1103/PhysRevD.83.024014}.
\newblock \urlprefix\url{https://link.aps.org/doi/10.1103/PhysRevD.83.024014}

\bibitem{PhysRevD.101.124006}
N.~Jiang, K.~Yagi, Analytic i-love-c relations for realistic neutron stars.
\newblock Phys. Rev. D \textbf{101}, 124006 (2020).
\newblock \doi{10.1103/PhysRevD.101.124006}.
\newblock \urlprefix\url{https://link.aps.org/doi/10.1103/PhysRevD.101.124006}

\bibitem{Breu_2016}
C.~Breu, L.~Rezzolla, {Maximum mass, moment of inertia and compactness of relativistic stars}.
\newblock Monthly Notices of the Royal Astronomical Society \textbf{459}(1), 646--656 (2016).
\newblock \doi{10.1093/mnras/stw575}.
\newblock \urlprefix\url{https://doi.org/10.1093/mnras/stw575}

\bibitem{PhysRevD.91.044034}
C.~Chirenti, G.H. de~Souza, W.~Kastaun, Fundamental oscillation modes of neutron stars: Validity of universal relations.
\newblock Phys. Rev. D \textbf{91}, 044034 (2015).
\newblock \doi{10.1103/PhysRevD.91.044034}.
\newblock \urlprefix\url{https://link.aps.org/doi/10.1103/PhysRevD.91.044034}

\bibitem{Yagi:2013bca}
K.~Yagi, N.~Yunes, {I-Love-Q}.
\newblock Science \textbf{341}, 365--368 (2013).
\newblock \doi{10.1126/science.1236462}.
\newblock {\href{https://arxiv.org/abs/1302.4499}{{arXiv:1302.4499}}} {[gr-qc]}

\bibitem{staykov2016}
K.V. Staykov, D.D. Doneva, S.S. Yazadjiev, Moment-of-inertia--compactness universal relations in scalar-tensor theories and ${\mathcal{r}}^{2}$ gravity.
\newblock Phys. Rev. D \textbf{93}, 084010 (2016).
\newblock \doi{10.1103/PhysRevD.93.084010}.
\newblock \urlprefix\url{https://link.aps.org/doi/10.1103/PhysRevD.93.084010}

\bibitem{Chan_2014}
T.K. Chan, Y.H. Sham, P.T. Leung, et~al., Multipolar universal relations between $f$-mode frequency and tidal deformability of compact stars.
\newblock Phys. Rev. D \textbf{90}, 124023 (2014).
\newblock \doi{10.1103/PhysRevD.90.124023}.
\newblock \urlprefix\url{https://link.aps.org/doi/10.1103/PhysRevD.90.124023}

\bibitem{Bikram_2023}
B.K. Pradhan, A.~Vijaykumar, D.~Chatterjee, Impact of updated multipole love numbers and $f$-love universal relations in the context of binary neutron stars.
\newblock Phys. Rev. D \textbf{107}, 023010 (2023).
\newblock \doi{10.1103/PhysRevD.107.023010}.
\newblock \urlprefix\url{https://link.aps.org/doi/10.1103/PhysRevD.107.023010}

\bibitem{GW190814}
R.~Abbott, T.D. Abbott, L.S. Collaboration, et~al., Gw190814: Gravitational waves from the coalescence of a 23 solar mass black hole with a 2.6 solar mass compact object.
\newblock The Astrophysical Journal Letters \textbf{896}(2), L44 (2020).
\newblock \doi{10.3847/2041-8213/ab960f}.
\newblock \urlprefix\url{https://dx.doi.org/10.3847/2041-8213/ab960f}

\bibitem{Roupas2021}
Z.~Roupas, Secondary component of gravitational-wave signal gw190814 as an anisotropic neutron star.
\newblock Astrophysics and Space Science \textbf{366}(1), 9 (2021).
\newblock \doi{10.1007/s10509-021-03919-5}.
\newblock \urlprefix\url{https://doi.org/10.1007/s10509-021-03919-5}

\bibitem{Kokkotas1999}
K.D. Kokkotas, B.G. Schmidt, Quasi-normal modes of stars and black holes.
\newblock Living Reviews in Relativity \textbf{2}(1), 2 (1999).
\newblock \doi{10.12942/lrr-1999-2}.
\newblock \urlprefix\url{https://doi.org/10.12942/lrr-1999-2}

\bibitem{Bharat_and_Landry}
B.~Kumar, P.~Landry, Inferring neutron star properties from gw170817 with universal relations.
\newblock Phys. Rev. D \textbf{99}, 123026 (2019).
\newblock \doi{10.1103/PhysRevD.99.123026}.
\newblock \urlprefix\url{https://link.aps.org/doi/10.1103/PhysRevD.99.123026}

\end{thebibliography}

\end{document}